\documentclass[pra,onecolumn,floatfix,a4paper,superscriptaddress]{revtex4}
\usepackage{bm,color,graphicx,amsmath,txfonts}

\usepackage[colorlinks, citecolor=blue,linkcolor=blue]{hyperref}

\usepackage{float}

\begin{document}
\title{Squeezed light and coherent feedback enhanced multiparameter quantum estimation in a double-cavity optomechanics: steady and dynamical states}

\author{Hamza Harraf}
\affiliation{LPHE-Modeling and Simulation, Faculty of Sciences, Mohammed V University in Rabat, Rabat, Morocco}	
\author{Mohamed Amazioug} 
\thanks{m.amazioug@uiz.ac.ma}
\affiliation{LPTHE-Department of Physics, Faculty of sciences, Ibnou Zohr University, Agadir, Morocco}

\author{Amjad Sohail}
\affiliation{Department of Physics, Government College University, Allama Iqbal Road,
Faisalabad 38000, Pakistan}

\author{Mojtaba Mazaheri}
\affiliation{Department of Basic Science, Hamedan University of Technology, Hamedan,
Iran}

\author{Rachid Ahl Laamara}
\affiliation{LPHE-Modeling and Simulation, Faculty of Sciences, Mohammed V University in Rabat, Rabat, Morocco}	
\affiliation{Centre of Physics and Mathematics, CPM, Faculty of Sciences, Mohammed V University in Rabat, Rabat, Morocco}

\begin{abstract}

Multiparameter quantum estimation in open optomechanical systems is fundamentally constrained by dissipation, thermal fluctuations, and measurement incompatibility. In this work, we investigate a coupled cavity optomechanical platform in which two mechanical modes interact with driven optical cavities, two mode squeezed vacuum, intracavity degenerate parametric amplification, and coherent optical feedback. Using the continuous variable Gaussian state formalism, we derive the linearized quantum Langevin dynamics and steady state covariance matrix and evaluate the quantum Fisher information matrices associated with simultaneous estimation of the optomechanical coupling strength and cavity dissipation rate. We characterize the precision bounds using the symmetric and right logarithmic derivative formalisms and employ $\mathcal{B}_{\rm MI}=\max\{\mathcal{B}_S,\mathcal{B}_R\}$ as a comparative figure of merit within the SLD/RLD framework. We find that parametric amplification can substantially reduce the most informative bound, $\mathcal{B}_{\rm MI}$, thereby enhancing multiparameter estimation precision over a broad range of operating conditions. Coherent feedback can further reduce $\mathcal{B}_{\rm MI}$ and improve the estimation precision, particularly in the strong feedback regime. Moreover, the simultaneous presence of coherent feedback, parametric amplification, and squeezed light injection provides a combined mechanism for controlling quantum fluctuations and parameter dependent correlations, leading to enhanced and more robust multiparameter estimation. We further analyze the dynamical and steady state regimes to identify the parameter regions in which these quantum resources provide the largest metrological gain. Our results establish coupled cavity optomechanics as a flexible platform for quantum-noise engineering in multiparameter sensing.

\end{abstract}

\maketitle

\section{Introduction}

Cavity optomechanics explores the radiation-pressure-mediated interplay between electromagnetic fields and mechanical resonators \cite{intro_14_5,intro_15,Asjad14,Agarwal13,Abdi121}. This light–matter coupling offers a powerful framework for generating and controlling nonclassical optical states, particularly squeezed light \cite{saheked2018,racz2023,frascella2021}. Over the last decade, such systems have gained significant interest due to their wide-ranging applications in quantum information science, including the creation of quantum entanglement \cite{Hofer11,weedbrook2012,Wang13,yan2019}, achieving ground-state cooling of mechanical oscillators \cite{intro_17,Abdi12,Asjad19,Agarwal09}, photon blockade \cite{Rabl11,intro_18}, integrating with hybrid superconducting quantum circuits \cite{intro_17}, and enabling coherent quantum control of macroscopic mechanical devices \cite{intro_20}.
Conceptually, cavity optomechanics extends cavity quantum electrodynamics (CQED), wherein photon-atom interactions are engineered within high-finesse optical or microwave resonators. In the strong-coupling limit of CQED, a single confined photon can profoundly alter the atom–cavity dynamics, giving rise to a host of striking quantum effects. Recent progress in reaching this regime has facilitated the demonstration of quantum phase gates \cite{intro_21}, the deterministic production of Fock states \cite{intro_22}, and quantum nondemolition measurements of individual cavity photons \cite{intro_23}. Continuous-variable (CV) Gaussian states \cite{Kim02,Paris05,Van05,Harraf26,Harraf262} play a central role in quantum information science due to their combination of mathematical simplicity and experimental practicality. Since these states are fully characterized by their first- and second-order statistical moments, they provide a highly efficient framework for studying multipartite quantum systems. Moreover, their generation, manipulation, and detection can be carried out with remarkable precision using existing experimental tools. Consequently, Gaussian states have been widely implemented in a range of quantum technologies, including cavity optomechanics \cite{Vitali07,Mauro07,Groblasher09,Tian10,intro_17,Chan11,Brunelli21,Amazioug2018,Amazioug2021,amazioug2020,Amazouig2020,harraf2026}, quantum teleportation \cite{Paris03}, and the emerging field of cavity magnomechanics \cite{Zhang16,Li18}. Their importance is further evidenced by their use in landmark experiments, such as Bose-Einstein condensates \cite{Kevrekidis03,Gross11,Wade16} and high-sensitivity gravitational-wave detectors \cite{Aasi13}.

At the core of quantum metrology is the quantum Fisher information matrix (QFIM), originally developed within the quantum estimation framework by Helstrom and Holevo \cite{Holevo11,Helstrom67,Helstrom76,Bures69} and subsequently formulated in a more general form by Braunstein and Caves \cite{Braunstein94}. The QFIM quantifies the sensitivity of a quantum state to infinitesimal variations of the unknown parameters and, therefore, determines the fundamental precision limits of multiparameter quantum estimation through the quantum Cram\'er--Rao bound (QCRB) \cite{paris2009}. In particular, the inverse QFIM provides a lower bound on the covariance matrix of any unbiased estimator, making the enhancement of quantum Fisher information a central objective in the development of high-precision quantum sensing protocols.

In contrast to the single-parameter case, where the QCRB can be asymptotically saturated under suitable conditions, its simultaneous saturation in multiparameter estimation is generally hindered by measurement incompatibility. This limitation originates from the noncommutativity of the optimal observables associated with different parameters, which can prevent the simultaneous realization of their individual optimal measurements \cite{Matsumoto02,Vaneph13,Vidrighin14,Crowley14,Ragy16,Szczykulska2016}. Nevertheless, the QCRB remains a fundamental benchmark for assessing the ultimate performance of quantum sensing strategies and for identifying quantum resources and operating regimes that provide an advantage over classical estimation schemes \cite{Giovannetti04,Zwierz10,Giovannetti11,Demkowicz12}.

The versatility of quantum metrology has motivated extensive investigations of both single- and multiparameter estimation in a wide range of physical systems. Representative applications include gravitational-wave detection \cite{Abbott16}, quantum thermometry \cite{Monras11,Hofer17,Correa15,Paris20,Khan22,Mirkhalaf24}, and the estimation of phases, squeezing parameters, time, and frequency \cite{Ballester04,Gaiba09,Zhang13,Frowis14}. Moreover, the QFIM is closely connected to fundamental quantum resources, including coherence and entanglement, establishing a direct link between quantum correlations and metrological performance \cite{Seveso19,Hauke16,Zhang13,Liu17,Binder22,FBMB23,Asjad2023,Cavazzoni25,Peng25}. In this context, Recent studies have highlighted the important role of multiparameter quantum estimation in characterizing and exploiting squeezed-light resources in cavity magnonic systems \cite{Hamza2026} and exciton--optomechanical systems \cite{Harraf2026}. These developments highlight the importance of investigating suitable quantum resources and control mechanisms for improving estimation precision, while also motivating the development of efficient theoretical frameworks for evaluating the QFIM and the associated precision bounds in increasingly complex quantum systems.\\

In this work, we theoretically investigate multiparameter quantum estimation in a feasible optomechanical system composed of two Fabry--P\'erot cavities connected through two-mode squeezed input fields and coherent-feedback loop implemented with a beam splitter. Each cavity contains a movable mechanical mirror, is driven by a coherent laser field, and incorporates a degenerate parametric amplifier (PA). The combination of coherent feedback and parametric amplification provides an experimentally accessible resource for enhancing the sensitivity of the system to the parameters of interest. We employ the quantum Fisher information matrix (QFIM) to characterize the ultimate precision limits of multiparameter estimation and focus on the most-informative quantum Cram\'er--Rao bound (QCRB), $\mathcal{B}_{\rm MI}$. We focus on the simultaneous estimation of the optomechanical coupling strength $g$ and cavity decay rate $\kappa_{\rm a}$, using the full QFIM, including its off-diagonal elements, and compare the corresponding SLD and RLD precision bounds. We investigate the dependence of the most informative bound, $\mathcal{B}_{\rm MI}$, on the main physical parameters of the system and examine the respective and combined effects of parametric amplification, coherent feedback, and squeezed light injection on the achievable multiparameter estimation precision. Our results show that parametric amplification can substantially reduce $\mathcal{B}_{\rm MI}$, thereby enhancing the estimation precision over a broad range of operating conditions. Coherent feedback can further reduce $\mathcal{B}_{\rm MI}$ and improve the estimation precision, particularly in the strong feedback regime. Moreover, the simultaneous presence of coherent feedback, parametric amplification, and squeezed light injection provides an effective means of controlling quantum fluctuations and parameter dependent correlations, leading to enhanced and more robust multiparameter estimation. Furthermore, we consider experimentally feasible Gaussian measurement strategies, namely homodyne and heterodyne detection, to reconstruct the first-order moments and covariance matrix of the output Gaussian state. In this way, we establish a direct connection between the ultimate quantum precision limits and experimentally accessible estimation strategies, demonstrating the potential of coherent feedback and parametric amplification for enhancing multiparameter quantum metrology.\\

This paper is organized as follows. In Sec. \ref{sec2}, we study the system dynamics using the quantum Langevin equations (QLEs). We also linearize these equations and derive the covariance matrix (CM) for the steady state. In Sec. \ref{sec4}, we present the theoretical tools for multiparameter quantum estimation of Gaussian states. Specifically, we calculate the QFIM using the SLD and RLD operators, compare single-parameter and simultaneous estimation, and compute the classical Fisher information for homodyne and heterodyne detection. Our numerical results are shown and discussed in Sec. \ref{sec5}. Finally, Sec. \ref{sec7} concludes the paper with a summary of our results and some perspectives for future work.

\section{Theoretical Model}\label{sec2}

\begin{figure}[!h]
\centering
\includegraphics[scale=0.45]{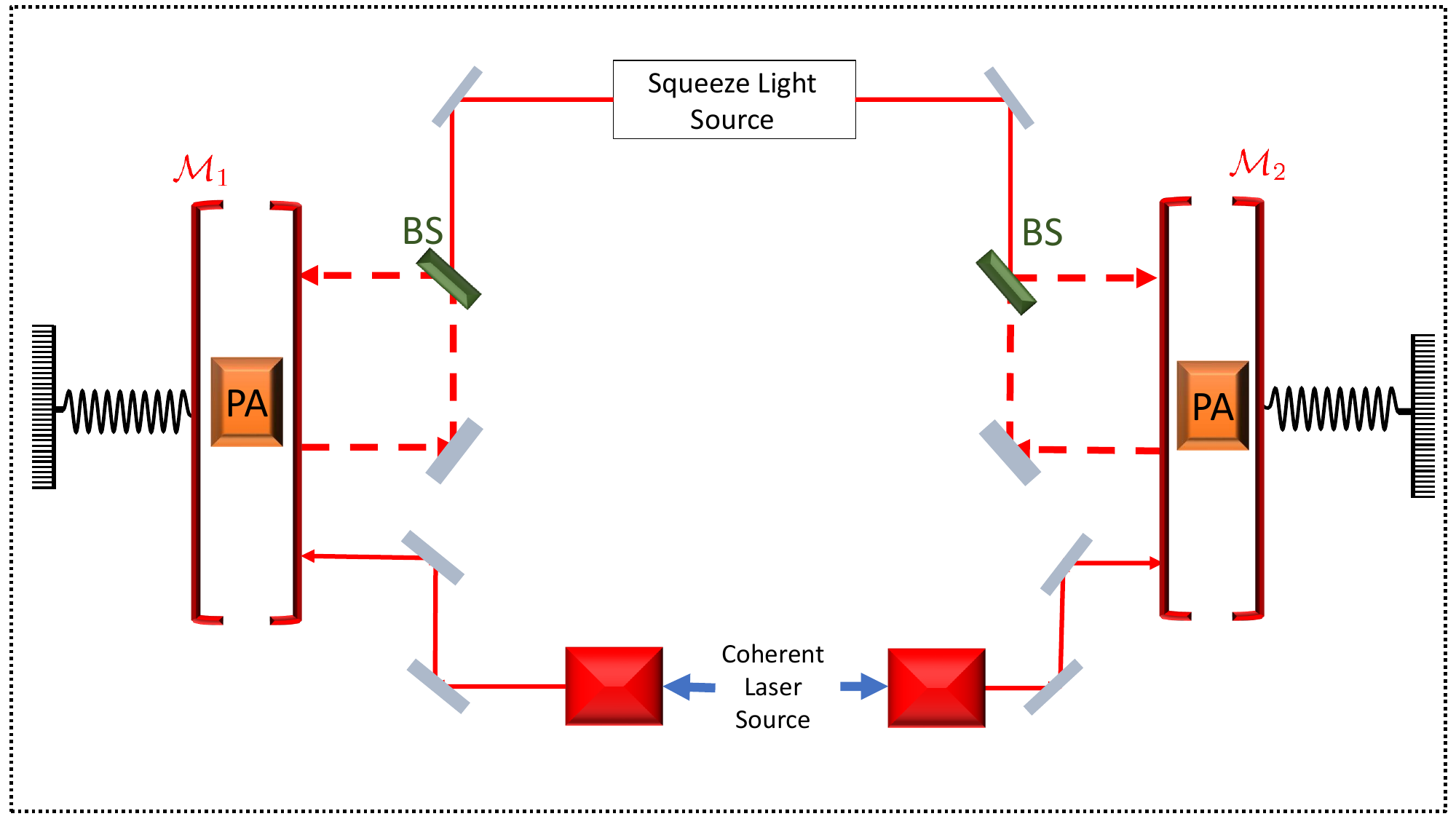} 
\caption{ Conceptual diagram of the proposed pair of symmetric Fabry--P\'erot optomechanical cavities incorporating coherent optical feedback. In each subsystem, the incident optical field is injected through an asymmetric beam splitter (BS), and the resulting output is reflected by an external mirror. A controlled portion of this reflected field is subsequently redirected through the BS and reinjected into the cavity, forming the coherent feedback path. The two cavities are independently driven by coherent laser fields of amplitudes $\mathrm{E}_i$ and receive the two correlated modes of a two-mode squeezed-vacuum field produced by spontaneous parametric down conversion. Each resonator is formed by a partially transmitting stationary mirror and a perfectly reflecting movable mirror $\mathcal{M}_i$ ($i\in\{1,2\}$), the latter providing the mechanical degree of freedom of each optomechanical subsystem. For simplicity of presentation, the pump fields associated with the intracavity parametric amplifiers are omitted from the schematic.}\label{fig_1}
\end{figure}

We consider the hybrid optomechanical configuration shown in Fig.~\ref{fig_1}, comprising two Fabry--P\'erot cavities driven by a common two-mode squeezed-vacuum field generated via parametric down-conversion. Each cavity consists of a partially transmitting fixed mirror and a perfectly reflecting movable mirror, with the latter acting as a mechanical oscillator. In addition to the coherent laser drive, each cavity is coupled to one mode of the two-mode squeezed-vacuum field, thereby establishing nonclassical correlations between the two optical subsystems. The movable mirrors, denoted by $\mathcal{M}_i$ ($i\in\{1,2\}$), are modeled as quantum harmonic oscillators and interact with their respective intracavity optical modes through radiation-pressure coupling. Throughout the analysis, we set $\hbar=1$. The Hamiltonian governing the coupled optomechanical system is then given by \cite{Sete2014,Amaziuog2020}
\begin{equation}
\label{0}
\mathcal{H}_{\text{total}}=\sum_{i=1}^{2}{}\left\{
\omega_{{\rm b}_i}{\rm b}_i^{\dagger}{\rm b}_i
+(\omega_{{\rm a}_i}-\omega_{{\rm L}_i}){\rm a}_i^{\dagger}{\rm a}_i
+{\rm g}_{i}{\rm a}_i^{\dagger}{\rm a}_i({\rm b}_i+{\rm b}_i^{\dagger}) 
+i\lambda_i({\rm a}_i^{\dagger 2}{\rm e}^{i\theta_i}{\rm e}^{-2i\omega_{{\rm b}_i}{\rm t}}
+{\rm a}_i^2{\rm e}^{-i\theta_i}{\rm e}^{2i\omega_{{\rm b}_i}{\rm t}})+\mu_{i}\left({\rm a}_i^{\dagger}\mathrm{E}_{i}{\rm e}^{i\phi_{i}}+{\rm a}_i\mathrm{E}_{i}{\rm e}^{i\phi_{i}}\right)\right\}.
\end{equation}
The two contributions appearing in Eq.~\eqref{0} can be specified as follows. The mechanical part consists of two independent harmonic oscillators, characterized by frequencies $\omega_{{\rm b}_i}$, effective masses ${\rm m}_i$, and the corresponding annihilation and creation operators ${\rm b}_i$ and ${\rm b}_i^\dagger$ ($i\in\{1,2\}$), which obey the standard bosonic commutation relations. Likewise, the two cavity modes are described by the frequencies $\omega_{{\rm a}_i}$ and the associated operators ${\rm a}_i$ and ${\rm a}_i^\dagger$. The optomechanical radiation-pressure interaction is characterized by the single-photon coupling strengths
$
{\rm g}_i=\frac{\omega_{{\rm a}_i}}{{\rm L}_i}\sqrt{\frac{\hbar}{{\rm m}_i\omega_{{\rm b}_i}}},
$
where ${\rm L}_i$ denotes the length of the $i$th cavity~\cite{intro_15}. Where the $i$th degenerate parametric amplifier (PA) is characterized by a gain parameter $\lambda_i$, which depends on both the PA pump strength and the phase $\theta_i$ of the field driving the amplifier. In this configuration, the pump field driving the PA has a frequency $2(\omega_{{\rm b}_i}+\omega_{{\rm L}_i})$ and interacts with a second-order nonlinear optical crystal. As a result, the signal and idler fields are degenerate, sharing the same frequency $\omega_{{\rm b}_i}+\omega_{{\rm L}_i}$~\cite{Agarwa2016}. The amplitude of the input field associated with the $i$th coherent laser source can be written as
$
\mathrm{E}_i=\sqrt{\frac{2\kappa_{{\rm a}_i}\mathcal{P}_i}{\hbar\omega_{{\rm L}_i}}}, \text{with } (i\in\{1,2\}),
$
where $\kappa_{{\rm a}_i}$ and $\mathcal{P}_i$ denote, respectively, the damping rate of the $i$th cavity and the corresponding driving power, while $\omega_{{\rm L}_i}$ is the frequency of the $i$th laser source. The coherent feedback is implemented through an asymmetric beam splitter, characterized by the real amplitude transmission and reflection coefficients $\mu_i$ and $\tau_i$, respectively. These coefficients satisfy
$
\mu_i^2+\tau_i^2=1,
$
with $\mu_i,\tau_i>0$~\cite{Mandel1995}. In the absence of coherent feedback, the output field of the $i$th cavity is not redirected into the same cavity, corresponding to the limiting case $\tau_i=0$ and $\mu_i=1$.
\\

The dynamics of the system, described by the non-linear quantum Langevin equations in a frame rotating at the driving laser frequency, can be expressed as follows \cite{Tian10}:
\begin{align}
\label{eq2}
\partial_{\rm t}{\rm b}_{i} &= -\left(i\omega_{{\rm b}_i} + \kappa_{{\rm b}_i}\right) {\rm b}_i - i g_i {\rm a}_i^\dagger {\rm a}_i + \sqrt{2\kappa_{{\rm b}_i}} {\rm b}_i^{\rm in},\\ \nonumber
\partial_{\rm t}{\rm a}_{i} &= -\left(\kappa_{{\rm fb}_i} - i\Delta_{{\rm fb}_i}\right) {\rm a}_i - i g_i {\rm a}_i ({\rm b}_i^\dagger + {\rm b}_i) +2\lambda_i {\rm e}^{i\theta_{i}}{\rm a}^{\dagger}_{i}{\rm e}^{-2i\omega_{{\rm b}_{i}}{\rm t}}-i\mu_{i}\mathrm{E}_{i}{\rm e}^{i\phi_{i}} + \sqrt{2\kappa_{{\rm a}_i}} {\rm a}_{\rm fb}^{\rm in},\nonumber
\end{align}
where $\{\partial_{\rm t}=\frac{d}{d{\rm t}}\}$ and $\kappa_{{\rm b}_i}$ denotes the dissipation rate of the $i$th movable mirror, while ${\rm b}_i^{\rm in}$ is the corresponding noise operator describing the interaction between the $i$th mechanical mode and its thermal environment. Within the Markovian approximation, the mechanical quality factor is assumed to be sufficiently large, i.e., $Q_i=\omega_{{\rm b}_i}/\kappa_{{\rm b}_i}\gg1$ ($i\in\{1,2\}$). Under this assumption, the mechanical noise operators satisfy the correlations
$
\langle {\rm b}_i^{\rm in}({\rm t}){\rm b}_i^{{\rm in}\dagger}({\rm t}')\rangle
=(n_{{\rm th}_i}+1)\delta({\rm t}-{\rm t}'),
$ and 
$
\langle {\rm b}_i^{{\rm in}\dagger}({\rm t}){\rm b}_i^{\rm in}({\rm t}')\rangle
=n_{{\rm th}_i}\delta({\rm t-t}'),
$
where
$
n_{{\rm th}_i}=
\left[\exp\left(\frac{\hbar\omega_{{\rm b}_i}}{k_{\rm B}{\rm T}_i}\right)-1\right]^{-1},
$
with $k_{\rm B}$ denoting the Boltzmann constant and ${\rm T}_i$ the temperature of the $i$th mechanical reservoir~\cite{Giovannetti2001,Gardiner2016}.

The coherent feedback modifies both the cavity decay rate and the effective detuning. In particular, the $i$th effective cavity decay rate and detuning are given, respectively, by
$
\kappa_{{\rm fb}_i}=\kappa_{{\rm a}_i}\left(1-2\tau_i\cos\alpha\right),
$
and
$
\Delta_{{\rm fb}_i}=\Delta_{{\rm a}_i}+\kappa_{{\rm a}_i}\tau_i\sin\alpha,
$
where $\Delta_{{\rm a}_i}=\omega_{{\rm a}_i}-\omega_{{\rm L}_i}$ and $\alpha$ denotes the phase shift acquired by the output field upon reflection from the feedback mirror.

The input optical field associated with the coherent feedback loop is described by the operator
$
{\rm A}_{{\rm fb}_i}^{\rm in}=\tau_i {\rm e}^{i\alpha}{\rm a}_i^{\rm out}+\mu_i {\rm a}_i^{\rm in},
$
which reflects the fact that the beam splitter transmits the fraction $\mu_i {\rm a}_i^{\rm in}$ of the incident field while reflecting the fraction $\tau_i {\rm a}_i^{\rm out}$ of the cavity output field. The reflected component is subsequently redirected into the cavity, thereby establishing the coherent feedback loop. The output field ${\rm a}_i^{\rm out}$ and the intracavity field ${\rm a}_i$ are related through the standard input--output relation
$
{\rm a}_i^{\rm out}=\sqrt{\kappa_{{\rm a}_i}}\,{\rm a}_i-\mu_i{\rm a}_i^{\rm in},
$ \cite{Walls1998}
so that the feedback input operator can equivalently be written as
$
{\rm A}_{{\rm fb}_i}^{\rm in}
=\tau_i\sqrt{\kappa_{{\rm a}_i}}{\rm e}^{i\alpha}{\rm a}_i
+{\rm a}_{{\rm fb}_i}^{{\rm in}},
$
where the effective optical input noise is defined as
$
{\rm a}_{{\rm fb}_i}^{{\rm in}}
=\mu_i\left(1-\tau_i {\rm e}^{i\alpha}\right){\rm a}_i^{\rm in}.
$
Here, ${\rm a}_i^{\rm in}$ represents the optical vacuum noise incident on the cavity through the beam splitter. Owing to the coherent feedback loop, the effective input noise ${\rm a}_{{\rm fb}_i}^{{\rm in}}$ exhibits nonzero correlations characteristic of a squeezed-vacuum field.
\begin{equation}
\langle {\rm a}^{\rm in}_{{\rm fb}_i}({\rm t}){\rm a}^{\rm in\dagger}_{{\rm fb}_i}({\rm t}')\rangle/\delta({\rm t}-{\rm t}')=\mu_i^2(1-\tau_i{\rm e}^{i\alpha})(1-\tau_i{\rm e}^{-i\alpha})(\mathrm{n}_i+1),\quad \langle {\rm a}^{\rm in\dagger}_{{\rm fb}_i}({\rm t}){\rm a}^{\rm in}_{{\rm fb}_i}({\rm t}')\rangle/\delta({\rm t}-{\rm t}')=\mu_i^2(1-\tau_i{\rm e}^{i\alpha})(1-\tau_i{\rm e}^{-i\alpha})\mathrm{n}_i,
\end{equation}
\begin{equation}
\langle {\rm a}^{\rm in}_{{\rm fb}_i}({\rm t}){\rm a}^{\rm in}_{{\rm fb}_i}({\rm t}')\rangle/\delta({\rm t}-{\rm t}')=\mu_i^2(1-\tau_i^2)(1-\tau_i{\rm e}^{i\alpha})(1-\tau_i{\rm e}^{-i\alpha})\mathrm{M}_i{\rm e}^{-i\omega_{{\rm b}_i}({\rm t}+{\rm t}')},\quad \langle {\rm a}^{\rm in\dagger}_{{\rm fb}_i}({\rm t}){\rm a}^{\rm in\dagger}_{{\rm fb}_i}({\rm t}')\rangle/\delta({\rm t}-{\rm t}')=\mu_i^2(1-\tau_i{\rm e}^{i\alpha})(1-\tau_i{\rm e}^{-i\alpha})\mathrm{M}_i{\rm e}^{i\omega_{{\rm b}_i}({\rm t}+{\rm t}')}.
\end{equation}
Where $\mathrm{n}_i=\sinh 2{\rm r_i}$ and $\mathrm{M}_i=\sinh {\rm r_i}\cosh {\rm r_i}$, with $\rm r_i$ denoting the squeezing parameter that characterizes the two-mode squeezed-vacuum field. In the present analysis, we consider identical mechanical frequencies for the two optomechanical subsystems, such that $\omega_{{\rm b}_1}=\omega_{{\rm b}_2}=\omega_{\rm b}$.

\subsection{Linearization of quantum Langevin equations}

The quantum Langevin equations in Eq.~(\ref{eq2}) are not generally amenable to exact analytical treatment. We therefore linearize them following the procedure detailed in Ref.~\cite{Sete2011} $ {\rm b}_i= \bar{\rm b}_i+\delta {\rm b}_i,$ ${\rm a}_i= \bar{\rm a} _i+\delta {\rm a}_i,$ ($i\in\{1,2\}$).
Where $\delta {\rm a}_i$ and $\delta {\rm b}_i$ are the operators of fluctuations; $\bar{\rm a}_i$ and
$\bar{\rm b}_i$ are the steady state averages for operators ${\rm a}_i$ and ${\rm b}_i$.
Utilizing Eq. \eqref{eq2}, one can get the clear expression of $\bar{\rm a}_{i}$ and $\bar{\rm b}_{i}$ :
\begin{equation}
\label{eq10}
 \bar{{\rm b}}_i= -\frac{i{\rm g}|\bar{\rm a}_i|}{i\omega_{{\rm b}_i}+\kappa_{{\rm b}_i}},\quad  \bar{\rm a}_{i}=-\frac{i\mu\mathrm{E}_i{\rm e}^{i\phi_i}}{\kappa_{{\rm fb}_i}-i\tilde{\Delta}_{{\rm fb}_i}}.
\end{equation}
Here, $\tilde{\Delta}_{{\rm fb}_i}=\Delta_{{\rm fb}_i}-{\rm g}_i\left(\bar{\rm b}_{i}+\bar{\rm b}_i^{*}\right)$, with $i\in\{1,2\}$, represents the effective cavity detuning, which accounts for the radiation-pressure-induced displacement of the corresponding movable mirror. Accordingly, Eq.~\eqref{eq2} can be recast in the following form:
\begin{align}
		\label{eq1_1}
		\partial_{\rm t}\delta {\rm b}_i&=-(\kappa_{{\rm b}_i}+i\omega_{{\rm b}_i})\delta {\rm b}_i+\mathcal{G}_i\left(\delta {\rm a}_i-\delta {\rm a}^{\dagger}_i\right)+\sqrt{2\kappa_{{\rm b}_i}}{\rm b}^{\rm in}_i,\\ \nonumber
		\partial_{\rm t}\delta {\rm a}_i&=-(\kappa_{{\rm fb}_i}-i\tilde{\Delta}_{{\rm fb}_i})\delta{\rm a}_{i}-\mathcal{G}_i\left(\delta{\rm b}_i+\delta{\rm b}^{\dagger}\right)+2\lambda_i{\rm e}^{i\theta_i}\delta{\rm a}^{\dagger}_i{\rm e}^{-2i\omega_{{\rm b}_i}{\rm t}}+\sqrt{2\kappa_{{\rm a}_i}}{\rm a}^{\rm in}_{{\rm fb}_i},\nonumber
\end{align}
where $\mathcal{G}_i={\rm g}_i|\bar{\rm a}_i|$  are the many-photon optomechanical coupling. We have selected the input field's phase $\phi_i$
to be $\phi_i=-\arctan\left(\frac{2\tilde{\Delta}_{{\rm a}_i}}{\kappa_{{\rm a}_i}}\right)$ and we have $\bar{\rm a}=-i|\bar{\rm a}_{i}|$. 
With introducing notation for operators  
$\delta {\rm b}_i({\rm t})=\delta \tilde{\rm b}_i({\rm t}){\rm e}^{i\omega_{{\rm b}_i}{\rm t}},$   $\delta {\rm a}_i({\rm t})=\delta \tilde{\rm a}_i({\rm t}){\rm e}^{i\tilde{\Delta}_{{\rm fb}_i}{\rm t}},
$
and 
$
	\begin{aligned}
		\tilde{\rm b}_i^{\rm in}={\rm e}^{i\omega_{{\rm b}_i}{\rm t}}{\rm b}^{\rm in}_i,&\quad \tilde{\rm a}_i^{\rm in}={\rm e}^{i\tilde{\Delta}_{{\rm fb}_i}{\rm t}}{\rm a}^{\rm in}_i.
	\end{aligned}
$

We apply the rotating-wave approximation (RWA) \cite{intro_15,Wang2015}, which is justified in the resolved-sideband regime, $\omega_{\rm b}\gg\kappa_{\rm a}$. In this regime, the rapidly oscillating counter-rotating terms at frequencies $\pm2\omega_{\rm b}$ can be safely neglected. For red-sideband driving, corresponding to the effective detuning $\tilde{\Delta}_{{\rm fb}_i}=-\omega_{\rm b}$, the interaction Hamiltonian reduces to the beam-splitter form, and Eq.~\eqref{eq1_1} becomes
\begin{equation}
	\begin{aligned}
		\label{eq_15}
		\partial_{\rm t}\delta\tilde{\rm b}_i=-\kappa_{{\rm b}_i}\delta\tilde{\rm b}_i+\mathcal{G}_i\delta\tilde{\rm a}_i+\sqrt{2\kappa_{{\rm b}_i}}\tilde{\rm b}^{\rm in}_i,&\quad
		\partial_{\rm t}\delta\tilde{\rm a}_i=-\kappa_{{\rm a}_i}\delta\tilde{\rm a}_i+\mathcal{G}_i\delta\tilde{\rm b}+2\lambda_i\cos\theta\delta\tilde{\rm a}_i^{\dagger}+2i\lambda_i\sin\theta\delta\tilde{\rm a}^{\dagger}_i+\sqrt{2\kappa_{{\rm a}_i}}\tilde{\rm a}^{\rm in}_i.
	\end{aligned}
\end{equation}

\subsection{Covariance matrix}

To characterize the precision of multiparameter quantum estimation in the proposed optomechanical system, we employ the covariance matrix (CM) formalism to describe the Gaussian state in both the dynamical and steady-state regimes. The CM provides a compact description of the quantum fluctuations and intermode correlations and constitutes a fundamental tool for evaluating the quantum Fisher information associated with the parameters of interest.

For simplicity, and without loss of generality, we consider the two subsystems to be identical and impose the following symmetric conditions:
$
n_{{\rm th}_1}=n_{{\rm th}_2}=n_{{\rm th}},$ $ {\rm T}_1={\rm T}_2={\rm T},
$ $
{\rm m}_1={\rm m}_2={\rm m},$ $
\omega_{{\rm a}_1}=\omega_{{\rm a}_2}=\omega_{\rm a},
$
$
\kappa_{{\rm fb}_1}=\kappa_{{\rm fb}_2}=\kappa_{\rm fb},
$ $
{\rm n}_1={\rm n}_2={\rm n},$
${\rm M}_1={\rm M}_2={\rm M},$
${\rm r}_1={\rm r}_2={\rm r},$
$\theta_1=\theta_2=\theta,$
$
\kappa_{{\rm b}_1}=\kappa_{{\rm b}_2}=\kappa_{\rm b},$ $
\lambda_1=\lambda_2=\lambda.
$
These symmetry conditions ensure that the two subsystems possess identical intrinsic properties and environmental parameters, thereby simplifying the analysis while preserving the essential quantum correlations induced by the two-mode squeezed-vacuum field.

Under these symmetric conditions, the linearized quantum dynamics of the two optomechanical subsystems can be conveniently described in terms of collective quadrature operators. This representation is particularly useful for characterizing the fluctuations and correlations shared by the optical and mechanical modes and for constructing the covariance matrix of the resulting Gaussian state. Since the CM contains the complete information required to characterize the Gaussian state, its dynamical and steady-state forms provide the basis for evaluating the sensitivity of the state to variations in the parameters to be estimated. Consequently, the corresponding collective quadratures are introduced to facilitate the derivation of the CM and its parameter derivatives, which are subsequently employed to evaluate the SLD and RLD-based quantum Fisher information matrices and the associated quantum Cram\'er--Rao bounds. The collective quadrature operators are defined as follows:
\begin{equation}
	\begin{aligned}
		\delta\tilde{P}_{{\rm b}_i}=\frac{\delta\tilde{\rm b}^{\dagger}_i+\delta\tilde{\rm b}_i}{\sqrt{2}},&\quad \quad \delta\tilde{Q}_{{\rm b}_i}=\frac{\delta\tilde{\rm b}_i-\delta\tilde{\rm b}^{\dagger}_i}{i\sqrt{2}}\\ 
		\delta\tilde{P}_{{\rm a}_i}=\frac{\delta\tilde{\rm a}^{\dagger}_i+\delta\tilde{\rm a}_i}{\sqrt{2}},&\quad \quad \delta\tilde{Q}_{{\rm a}_i}=\frac{\delta\tilde{\rm a}_i-\delta\tilde{\rm a}^{\dagger}_i}{i\sqrt{2}}.\\ 
	\end{aligned}
\end{equation}
This allows Eq. \eqref{eq_15} to be expressed compactly in the basis of the associated quadrature operators
\begin{equation}
\label{eq_18}
\begin{aligned}
\partial_{\rm t}\delta\tilde{P}_{{\rm b}_i}
={}&-\kappa_{{\rm b}_i}\delta\tilde{P}_{{\rm b}_i}
+\mathcal{G}_i\delta\tilde{Q}_{{\rm a}_i}
+\sqrt{2\kappa_{{\rm b}_i}}\tilde{Q}^{\rm in}_{{\rm b}_i},
\\
\partial_{\rm t}\delta\tilde{Q}_{{\rm b}_i}
={}&-\kappa_{{\rm b}_i}\delta\tilde{P}_{{\rm b}_i}
+\mathcal{G}_i\delta\tilde{Q}_{{\rm a}_i}
+\sqrt{2\kappa_{{\rm b}_i}}\tilde{Q}^{\rm in}_{{\rm b}_i}.
\end{aligned}
\end{equation}
and
\begin{equation}
\begin{aligned}
\partial_{\rm t}\delta\tilde{P}_{{\rm a}_i}
={}&-\kappa_{{\rm fb}_i}\delta\tilde{Q}_{{\rm a}_i}
+\mathcal{G}_i\delta\tilde{Q}_{{\rm b}_i}
+2\lambda_i\cos(\theta_i)\delta\tilde{P}_{{\rm a}_i}
+2\lambda_i\sin(\theta_i)\delta\tilde{Q}_{{\rm a}_i}
+\sqrt{2\kappa_{{\rm a}_i}}\tilde{Q}^{\rm in}_{{\rm a}_i},
\\
\partial_{\rm t}\delta\tilde{Q}_{{\rm a}_i}
={}&-\kappa_{{\rm fb}_i}\delta\tilde{Q}_{{\rm a}_i}
+\mathcal{G}_i\delta\tilde{Q}_{{\rm b}_i}
-2\lambda_i\cos(\theta_i)\delta\tilde{P}_{{\rm a}_i}
+2\lambda_i\sin(\theta_i)\delta\tilde{Q}_{{\rm a}_i}
+\sqrt{2\kappa_{{\rm a}_i}}\tilde{Q}^{\rm in}_{{\rm a}_i}.
\end{aligned}
\end{equation}
With
\begin{equation}
	\begin{aligned}
		\tilde{P}_{{\rm b}_i}^{\rm in}=\frac{\tilde{\rm b}^{\rm in\dagger}_i+\tilde{\rm b}^{\rm in}_i}{\sqrt{2}},&\quad \quad \quad \tilde{Q}_{{\rm b}_i}^{\rm in}=\frac{\tilde{\rm b}_i^{\rm in}-\tilde{\rm b}^{\rm in\dagger}_i}{i\sqrt{2}},\\ 
		\tilde{P}_{{\rm a}_i}^{\rm in}=\frac{\tilde{\rm a}^{\rm in\dagger}_i+\tilde{\rm a}^{\rm in}_i}{\sqrt{2}},&\quad \quad \quad \tilde{Q}_{{\rm a}_i}^{\rm in}=\frac{\tilde{\rm a}_i^{\rm in}-\tilde{\rm a}^{\rm in\dagger}_i}{i\sqrt{2}}.\\ 
	\end{aligned}
\end{equation}
Eqs. \eqref{eq_18} can be written in a compact matrix form \cite{Mari2009}:
\begin{equation}
\partial_{\rm t}\mathrm{u}({\rm t})=\mathcal{S}\mathrm{u}({\rm t})+\mathrm{v}({\rm t})
\end{equation}
with \begin{align}
\mathrm{u}({\rm t})=\left[\delta \tilde{P}_{{\rm b}_i}({\rm t}),\delta \tilde{Q}_{{\rm b}_i}({\rm t}),\delta \tilde{P}_{{\rm a}_i}({\rm t}),\delta \tilde{Q}_{{\rm a}_i}({\rm t}))\right]^T,&\quad\quad
\mathrm{v}({\rm t})=\left[ \tilde{P}^{\rm in}_{{\rm b}_i}({\rm t}), \tilde{Q}^{\rm in}_{{\rm b}_i}({\rm t}), \tilde{P}^{\rm in}_{{\rm a}_i}({\rm t}), \tilde{Q}^{\rm in}_{{\rm a}_i}({\rm t})\right]^T, \nonumber
\end{align}
and the corresponding drift matrix $\mathcal{S}$ takes the form:
\begin{equation}
\mathcal{S}=
\begin{pmatrix}
-\kappa_{\rm b}\,\mathbb{I}_4 & \mathcal{G}\,\mathbb{I}_4 \\[4pt]
-\mathcal{G}\,\mathbb{I}_4 & \mathcal{S}_{\rm fb}
\end{pmatrix},
\end{equation}
where $\mathbb{I}_4$ is the $4\times4$ identity matrix, and
$
\mathcal{S}_{\rm fb} = -\kappa_{\rm fb}\,\mathbb{I}_4
+ 2\lambda\,\mathbb{I}_2\otimes
\begin{pmatrix}
\cos\theta & \sin\theta \\
\sin\theta & -\cos\theta
\end{pmatrix}.
$\\
The system is dynamically stable provided that all eigenvalues of the drift matrix $\mathcal{S}$ possess negative real parts. This stability condition is satisfied for $\mathcal{G}>\kappa_{\rm a},\kappa_{\rm b}$, or equivalently $2\mathcal{G}>\kappa_{\rm a}+\kappa_{\rm b}$ \cite{DeJesus1987}. Under this condition, the steady state of the bipartite Gaussian system is completely characterized by the covariance matrix, whose elements are defined as
$
\mathcal{C}_{ij}
=\frac{\left\langle
\mathrm{u}_{i}({\rm t})\mathrm{u}_{j}({\rm t}')
+\mathrm{u}_{i}({\rm t}')\mathrm{u}_{j}({\rm t})
\right\rangle}{2}.
$
The dynamic-state covariance matrix is then obtained by solving the Lyapunov equation \cite{Vitali07}:
\begin{equation}
\partial_{\rm t}\mathcal{C}=\mathcal{S}\mathcal{C}+\mathcal{C} \mathcal{S}^T+\mathbb{F},
\end{equation}
where $\mathbb{F}$ is the matrix of stationary noise correlations functions given by $\mathbb{F}_{ij}=\frac{\langle \mathrm{v}_{i}({\rm t})\mathrm{v}_{j}({\rm t}')+\mathrm{v}_{i}({\rm t}')\mathrm{v}_{j}({\rm t})\rangle}{2\delta({\rm t}-{\rm t}')}$.
\begin{equation}
\mathbb{F} = 
\begin{pmatrix}
{\rm K}_{\rm b}\,\mathbb{I}_4 & 0 \\[4pt]
0 & \mathbb{F}_{\rm fb}
\end{pmatrix},
\end{equation}
where
$
\mathbb{F}_{\rm fb} = {\rm K}_{\rm fb}\,\mathbb{I}_4 
+ \mathrm{M}_{\rm fb}\,(\sigma_x\otimes\sigma_z),
$
and
 $\mathbb{I}_4$ is the $4\times4$ identity matrix, and
$
\sigma_x=\begin{pmatrix}0&1\\1&0\end{pmatrix},$ $
\sigma_z=\begin{pmatrix}1&0\\0&-1\end{pmatrix}
$
are the usual Pauli matrices, with
$
{\rm K}_{\rm b}=\kappa_{\rm b}(2n+1),$ 
$
{\rm K}_{\rm fb}=\kappa_{\rm a}(2n+1)\mu^2(1-\tau^2),$
$
\mathrm{M}_{\rm fb}=\mathrm{M}\,\kappa_{\rm a}\,\mu^2(1-\tau)^2.
$

\section{\label{sec4}QUANTUM AND CLASSICAL FISHER INFORMATION}

In this section, we first review the fundamental concepts and mathematical definitions underlying multiparameter quantum estimation, with particular emphasis on the characterization of its ultimate precision limits. We then formulate the corresponding framework for continuous-variable Gaussian quantum systems, which provides a convenient and efficient description of the quantum states considered in our analysis.

\subsection{Quantum multiparameter estimation \label{sec4.1}}

Multiparameter quantum estimation is a direct extension of the traditional single-parameter quantum metrology approach to the estimation of multiple physical parameters simultaneously. In the typical scheme of an estimation procedure, the quantum system under consideration is first prepared in a certain state, then evolves through a quantum process depending on the set of parameters of interest, and finally subjected to a quantum measurement providing some knowledge about the encoded parameters. In our study, we focus on a set of quantum states, denoted by $\rho_{\boldsymbol{\varrho}}$, in which $\boldsymbol{\varrho}=\{\varrho_{\epsilon}\}$ with ($\epsilon=1\cdots \rm k$) is the vector of $\rm k$ unknown parameters of interest. The estimation procedure consists of a quantum measurement characterized by a positive operator-valued measure (POVM), ${\Xi_{\mathbf{z}}}$, and an estimator $\hat{\boldsymbol{\varrho}}(\mathbf{z})$, which assigns a value
$
\sum_{\mathbf{z}}\Xi_{\mathbf{z}}^{\dagger}\Xi_{\mathbf{z}}=\mathbb{I},
$
while the conditional probability of obtaining the outcome $\mathbf{z}$ is given by the Born rule,
$
\mathrm{P}(\mathbf{z}|\boldsymbol{\varrho})
=\mathrm{Tr}\!\left[\rho_{\boldsymbol{\varrho}}\Xi_{\mathbf{z}}\right].
$
The ultimate precision achievable in estimating the parameters is characterized by the logarithmic derivative operators associated with each parameter $\rho_{\varrho}$. In particular, the Symmetric Logarithmic Derivative (SLD) and the Right Logarithmic Derivative (RLD) are defined, respectively, by \cite{EPJD2014,hmz:9,Safranek2018,Bakmou2020,Razhin2017,hmz:27}
\begin{align}
\frac{\partial\hat{\rho}}{\partial\boldsymbol{\varrho}_{\epsilon}}  &= \frac{\left\{\hat{\rho} ,\mathcal{L}^{(\rm S)}_{\varrho_{\epsilon}}\right\}}{2} \quad {(\rm SLD)}.  \label{eq:SLD} \\
\frac{\partial\hat{\rho}}{\partial\boldsymbol{\varrho}_{\epsilon}} &=\hat{\rho} {\mathcal{L}}_{\varrho_{\epsilon}}^{(\rm R)}  \quad {(\rm RLD)}. \label{eq:RLD}
\end{align}
 Within the SLD and RLD formalisms, the corresponding quantum Fisher information matrices (QFIMs) are defined as follows:
\begin{align}
{\mathbb{H}}_{\varrho_{\epsilon}\varrho_{\varepsilon}} &=  \frac{1}{2}{\rm Tr} \left[ \hat{\rho}\left\{{\mathcal{L}}^{\rm (S)}_{\varrho_{\epsilon}},{\mathcal{L}}^{\rm (S)}_{\varrho_{\varepsilon}} \right\}\right ]\:,  \\
{\mathbb{J}}_{\varrho_{\epsilon}\varrho_{\varepsilon}} &= {\rm Tr}[ \hat{\rho} {\mathcal{L}}_{\varrho_{\epsilon}}^{(\rm R)} {\mathcal{L}}_{\varrho_{\varepsilon}}^{(\rm R)\dag}],
\end{align}
where ${\rm Tr}[\alpha]$ denotes the trace of the finite-dimensional matrix $\alpha$. The attainable precision in multiparameter estimation is subject to two distinct Cram\'er--Rao-type inequalities. Specifically, by defining the covariance matrix of the measurement outcomes as
$
{\rm \bf Cov}(\boldsymbol{\hat{\varrho}})_{\epsilon\varepsilon}
=
{\rm \bf E}\!\left[\hat{\varrho}_{\epsilon}\hat{\varrho}_{\varepsilon}\right]
-
{\rm \bf E}\!\left[\hat{\varrho}_{\epsilon}\right]
{\rm \bf E}\!\left[\hat{\varrho}_{\varepsilon}\right],
$
where ${\rm \bf E}[\cdot]$ denotes the statistical expectation value \cite{hmz:27,Bakmou2020}, we obtain
\begin{align}
\textbf{Cov}(\hat{\boldsymbol{\varrho}})
&\geq
\frac{1}{\mathbb{N}}
\mathbb{H}^{-1},
\label{M}
\\
\textbf{Cov}(\hat{\boldsymbol{\varrho}})
&\geq
\frac{1}{\mathbb{N}}
\left[
\operatorname{Re}\!\left(\mathbb{J}^{-1}\right)
+
\left|
\operatorname{Im}\!\left(\mathbb{J}^{-1}\right)
\right|
\right],
\label{S}
\end{align}
where the inverse of the Quantum Fisher Information Matrix is decomposed into its real and imaginary parts,
$
\mathbb{J}^{-1}
=
\operatorname{Re}\!\left(\mathbb{J}^{-1}\right)
+
i\,\operatorname{Im}\!\left(\mathbb{J}^{-1}\right).
\label{eq:J_inverse_decomposition}
$
We further introduce the operator absolute value, defined as
$
|\alpha|\equiv\sqrt{\alpha\alpha^{\dagger}},
$
and let $\mathbb{N}$ denote the number of independent measurement repetitions i.e., the sample size.
Under these definitions, the inequalities in question provide lower bounds on the total variance of the estimated parameters.

In the special case of individual estimation, the off-diagonal terms vanish identically, such that
$
\mathbb{H}_{\varrho_{\epsilon}\varrho_{\varepsilon}} = \mathbb{J}_{\varrho_{\epsilon}\varrho_{\varepsilon}} = 0,  \text{for } \epsilon \neq \varepsilon.
$
Consequently, each parameter is characterized independently by its own variance, and Eqs.~(\ref{M}) and (\ref{S}) reduce to
\begin{align}
\mathrm{Var}\!\left[\hat{\varrho}_{\epsilon}\right]
&\geq
\frac{
\mathbb{H}^{-1}_{\varrho_{\epsilon}\varrho_{\epsilon}}
}{\mathbb{N}},
\label{M1}
\\
\mathrm{Var}\!\left[\hat{\varrho}_{\epsilon}\right]
&\geq
\frac{
\operatorname{Re}\!\left(
\mathbb{J}^{-1}_{\varrho_{\epsilon}\varrho_{\epsilon}}
\right)
+
\left|
\operatorname{Im}\!\left(
\mathbb{J}^{-1}_{\varrho_{\epsilon}\varrho_{\epsilon}}
\right)
\right|
}{\mathbb{N}}.
\label{S1}
\end{align}
Eqs. (\ref{M1}) and (\ref{S1}) are always saturable, marking the point of optimal parameter estimation. The optimal states are obtained via projections onto the SLD eigenbasis. Additionally, applying the trace operator to the inequalities in (\ref{M}) and (\ref{S}) shows that these expressions simplify to the total variance of the estimated parameters.
\begin{align}
\sum_{\alpha=1}^{\rm k}
\mathrm{Var}\!\left[\hat{\varrho}_{\epsilon_\alpha}\right]
&\geq
\frac{\mathcal{B}_{\mathrm{S}}}{\mathbb{N}}
:=
\frac{\operatorname{Tr}\!\left[\mathbb{H}^{-1}\right]}
{\mathbb{N}},
\label{BS}
\\
\sum_{\alpha=1}^{\rm k}
\mathrm{Var}\!\left[\hat{\varrho}_{\epsilon_\alpha}\right]
&\geq
\frac{\mathcal{B}_{\mathrm{R}}}{\mathbb{N}}
:=
\frac{
\operatorname{Tr}\!\left[
\operatorname{Re}\!\left(\mathbb{J}^{-1}\right)
\right]
+
\operatorname{Tr}\!\left[
\left|
\operatorname{Im}\!\left(\mathbb{J}^{-1}\right)
\right|
\right]
}
{\mathbb{N}}.
\label{RD}
\end{align}
The matrices $\mathbb{H}$ and $\mathbb{J}$ denote the SLD- and RLD-based quantum Fisher information matrices, respectively \cite{hmz:11,hmz:13,hmz:14}. In multiparameter quantum estimation, the attainability of the associated SLD and RLD Cram\'er--Rao bounds is generally governed by compatibility conditions between the parameters \cite{hmz:15}. In particular, the noncommutativity of the optimal observables associated with different parameters may prevent their individual optimal measurements from being simultaneously realized by a common measurement strategy. Consequently, optimizing the estimation of one parameter can affect the precision achievable for another. In addition, although the RLD formalism provides a useful bound on the attainable precision, its optimal estimator does not necessarily correspond to a physically realizable positive-operator-valued measure (POVM). Importantly, however, the noncommutativity of the optimal measurements for individual parameters does not by itself rule out the simultaneous attainability of the corresponding precision limits, since suitably designed collective or joint measurements may overcome the incompatibility at the level of separate measurements.

Although the SLD-based formulation has received considerable attention in quantum multiparameter estimation \cite{hmz:16,hmz:17,hmz:18,hmz:19,Monras11,hmz:21,Crowley14}, it is well known that, in the single-parameter scenario, the SLD quantum Fisher information generally does not exceed the corresponding RLD quantum Fisher information. Consequently, the SLD formulation provides the tighter sensitivity bound in this regime \cite{hmz:9,hmz:10}. Furthermore, while the single-parameter SLD Cram\'er--Rao bound is asymptotically attainable in the limit of a large number of measurements $\mathbb{N}$, its attainability in the multiparameter case requires additional compatibility conditions. In particular, recent results based on quantum local asymptotic normality indicate that the bound in Eq.~(\ref{M}) is asymptotically saturable when the weak compatibility condition is fulfilled \cite{hmz:16}, namely,
$
\operatorname{Tr}\!\left[
\hat{\rho}
\left[
\mathcal{L}_{\varrho_\epsilon},
\mathcal{L}_{\varrho_\varepsilon}
\right]
\right]=0.
$
When this condition is not satisfied, the simultaneous attainability of the SLD bound is generally obstructed by measurement incompatibility. In such cases, the RLD bound can provide an alternative and potentially more restrictive characterization of the ultimate precision achievable in the multiparameter estimation of the system parameters.

This naturally raises the question of when these inequalities are saturated in multiparameter settings, as this would enable the realization of an optimal measurement strategy. In this regard, it is instructive to note that while the SLD-based framework has been the subject of numerous investigations in quantum multiparameter estimation \cite{Ragy16, hmz:24, Vaneph13, Vidrighin14}, these works generally establish that the QCRBs in Eqs.~(\ref{M}) and (\ref{BS}) are saturable if and only if:
\begin{equation} 
{\rm Tr}\left[ \varrho_\epsilon \left[ \mathcal{L}_{\varrho_\epsilon}^{(S)}, \mathcal{L}_{\varrho_\epsilon}^{(S)} \right] \right] = 0. \label{23}
 \end{equation} 
 It is straightforward to show that condition~(\ref{23}) can be reformulated as
$
{\rm{Im}}\left( {{\rm Tr}\left[ { \hat{\rho} {\mathcal{L}}_{{\varrho_\epsilon}}^{(S)} {\mathcal{L}}_{{\varrho_\varepsilon}}^{(S)}} \right]} \right) = 0. \label{24}
$ It is consequently of considerable interest to elucidate the connection between the RLD and SLD bounds, with a view to establishing their respective foundational status. This pursuit ultimately gave rise to the most informative QCRB ($\mathcal{B}_{\rm MI}$) \cite{hmz:27, hmz:28}, which is defined as follows:
\begin{equation}
{\mathcal{B}_{\rm MI}} = \max\left\{ {{\mathcal{B}_{\rm R}},{\mathcal{B}_{\rm S}}} \right\}. \label{25}
\end{equation}
To identify the tightest precision limit within the two quantum estimation formalisms, we compare the bounds obtained from the SLD and RLD quantum Fisher information matrices. The most informative quantum Cram\'er--Rao bound is then selected according to their relative magnitude. To quantify the difference between these two bounds, we define the ratio

\begin{equation}
\mathcal{R} = \frac{{{\mathcal{B}_{\rm S}}}}{{{\mathcal{B}_{\rm R}}}}.
\end{equation}
The value of $\mathcal{B}_{\rm MI}$ is determined by the ratio $\mathcal{R}$ as follows:
\begin{equation}
\mathcal{B}_{\rm MI} = 
\begin{cases} 
\mathcal{B}_{\rm S}, & \mathcal{R} < 1, \\
\mathcal{B}_{\rm R}, & \mathcal{R} > 1,
\end{cases}
\end{equation}
with $\mathcal{B}_{\rm MI} = \mathcal{B}_{\rm R} = \mathcal{B}_{\rm S}$ for $\mathcal{R} = 1$.

Thus, the multiparameter precision limits reduce to the single inequality
\begin{equation}
\sum_{\epsilon} \operatorname{Var}[\varrho_\epsilon] \ge \frac{\mathcal{B}_{\rm MI}}{\mathbb{N}}.
\end{equation}
The SLD- and RLD-based quantum Fisher information matrices are derived in detail in Refs.~\cite{hmz:28,Bakmou2020}. In particular, when the matrix $\sigma=2\mathcal{C}+i\Omega$ is invertible, where $\mathcal{C}$ denotes the covariance matrix and
$
\Omega=
\begin{bmatrix}
0 & 1\\
-1 & 0
\end{bmatrix}
\oplus
\begin{bmatrix}
0 & 1\\
-1 & 0
\end{bmatrix}
$, is the symplectic form, the RLD quantum Fisher information matrix can be written as
\begin{equation}
{{\mathbb{J}}_{{\varrho_\epsilon}{\varrho_\varepsilon}}} = 2\mathtt{vec}{\left[ {{\partial _{{\varrho_\epsilon}}}\mathcal{C}} \right]^\dag }{\Sigma ^{ - 1}}\mathtt{vec}\left[ {{\partial _{{\varrho_\varepsilon}}}\mathcal{C} } \right] + 2{\partial _{{\varrho_{\epsilon} }}}{\mathbf{r}^\top}\hspace{0.1cm}\sigma ^{-1}\hspace{0.1cm}{\partial _{{\varrho _{\varepsilon} }}}\mathbf{r}. \label{RLD}
\end{equation}
Under this condition, the Moore--Penrose pseudoinverse of $\sigma$ coincides with its ordinary inverse. We then define
$
\Sigma=\sigma^{\dagger}\otimes\sigma,
$
while the displacement vector is given by
$
\mathbf{r}
=
\left[
\langle\hat{P}_{{\rm b}_i}\rangle,
\langle\hat{Q}_{{\rm b}_i}\rangle,
\langle\hat{P}_{{\rm a}_i}\rangle,
\langle\hat{Q}_{{\rm a}_i}\rangle
\right]^{\top}.
$
Here, $\mathtt{vec}[\cdot]$ denotes the vectorization operation that transforms a matrix into a column vector \cite{Safranek2018}. The elements of the SLD-based quantum Fisher information matrix can then be expressed as \cite{EPJD2014,hmz:9,Bakmou2020}:
\begin{equation}
{\mathbb{H}_{{\varrho_\epsilon }{\varrho_\varepsilon }}} = 2\mathtt{vec}{\left[ {{\partial _{{\varrho_\epsilon}}}\mathcal{C} } \right]^\dag }{{\mathcal Z}^{+}}\mathtt{vec}\left[ {{\partial _{{\varrho_\varepsilon}}}\mathcal{C} } \right] + {\partial _{{\varrho_\epsilon}}}{{\bf{r}}^\top} \hspace{0.1cm} \mathcal{C}^{-1} \hspace{0.1cm}{\partial _{{\varrho_\varepsilon}}}{\bf{r}},\label{37}
\end{equation}
with $\mathcal{Z} = 4{\mathcal{C}^\dagger \otimes \mathcal{C} + \Omega \otimes \Omega}$. Provided that $\mathcal{Z}$ is invertible, the SLD quantum Fisher information matrix can be expressed as
\begin{equation}
\label{SLD}
{\mathbb{H}_{{\varrho_\epsilon }{\varrho_\varepsilon }}} = 2\mathtt{vec}{\left[ {{\partial _{{\varrho_\epsilon}}}\mathcal{C} } \right]^\dag }{{\mathcal Z}^{-1}}\mathtt{vec}\left[ {{\partial _{{\varrho_\varepsilon}}}\mathcal{C} } \right] + {\partial _{{\varrho_\epsilon}}}{{\bf{r}}^\top} \hspace{0.1cm} \mathcal{C}^{-1} \hspace{0.1cm}{\partial _{{\varrho_\varepsilon}}}{\bf{r}},
\end{equation}
The SLD and RLD quantum Fisher information matrices are derived in detail in \cite{EPJD2014,hmz:9,Safranek2018,Bakmou2020}. The estimation of the
parameters: the cavity-phonon coupling $\rm g$ or dissipation rate of cavity mode $\kappa_{\rm a}$, the RLD quantum Fisher information matrix
$\mathbb{J}$, can be calculated as
\begin{eqnarray}\label{1}
\mathbb{J}&=&\begin{pmatrix}
2\mathtt{vec}{\left[ {{\partial _{{\rm g}}}\mathcal{C}} \right]^\dag }{\Sigma ^{ - 1}}\mathtt{vec}\left[ {{\partial_{{\rm g}}}\mathcal{C} } \right] + 2{\partial _{{\rm g}}}{\mathbf{r}^\top}\hspace{0.1cm}\sigma ^{-1}\hspace{0.1cm}{\partial _{{\rm g}}}\mathbf{r} & 2\mathtt{vec}{\left[ {{\partial _{{\rm g}}}\mathcal{C}} \right]^\dag }{\Sigma^{ - 1}}\mathtt{vec}\left[ {{\partial_{{\rm \kappa_{\rm a}}}}\mathcal{C} } \right] + 2{\partial _{{\rm g}}}{\mathbf{r}^\top}\hspace{0.1cm}\sigma ^{-1}\hspace{0.1cm}{\partial _{{\rm \kappa_{a}}}}\mathbf{r} \\
2\mathtt{vec}{\left[ {{\partial _{{\rm \kappa_{a}}}}\mathcal{C}} \right]^\dag }{\Sigma ^{ - 1}}\mathtt{vec}\left[ {{\partial_{{\rm g}}}\mathcal{C} } \right] + 2{\partial _{{\rm \kappa_{\rm a}}}}{\mathbf{r}^\top}\hspace{0.1cm}\sigma ^{-1}\hspace{0.1cm}{\partial _{{\rm g}}}\mathbf{r} & 2\mathtt{vec}{\left[ {{\partial _{\kappa_{\rm a}}}\mathcal{C}} \right]^\dag }{\Sigma ^{ - 1}}\mathtt{vec}\left[ {{\partial_{\kappa_{\rm a}}}\mathcal{C} } \right] + 2{\partial _{{\rm \kappa_{a}}}}{\mathbf{r}^\top}\hspace{0.1cm}\sigma ^{-1}\hspace{0.1cm}{\partial _{{\rm \kappa_{a}}}}\mathbf{r}
\end{pmatrix}, \label{RLDM}
\nonumber\\
&&=\begin{pmatrix}
{\mathbb{J}}_{{\rm g}{\rm g}} & {\mathbb{J}}_{\rm g\kappa_{\rm a}} \\
{\mathbb{J}}_{\kappa_{\rm a}{\rm g}} & {\mathbb{J}}_{\kappa_{\rm a}\kappa_{\rm a}}
\end{pmatrix},\
\end{eqnarray}
the SLD quantum Fisher information matrix $\mathbb{H}$, can be calculated from Eq. (\ref{SLD}), as
\begin{eqnarray}\label{SLDM}
\mathbb{H}&=& \begin{pmatrix}
2\mathtt{vec}{\left[ {{\partial _{{\rm g}}}\mathcal{C}} \right]^\dag }{\mathcal{Z} ^{ - 1}}\mathtt{vec}\left[ {{\partial_{{\rm g}}}\mathcal{C} } \right] + {\partial _{{\rm g}}}{\mathbf{r}^\top}\hspace{0.1cm}\mathcal{C}^{-1}\hspace{0.1cm}{\partial _{{\rm g}}}\mathbf{r} & 2\mathtt{vec}{\left[ {{\partial _{{\rm g}}}\mathcal{C}} \right]^\dag }{\mathcal{Z}^{ - 1}}\mathtt{vec}\left[ {{\partial_{{\rm \kappa_{a}}}}\mathcal{C} } \right] + {\partial _{{\rm g}}}{\mathbf{r}^\top}\hspace{0.1cm}\mathcal{C}^{-1}\hspace{0.1cm}{\partial _{{\rm \kappa_{a}}}}\mathbf{r} \\
2\mathtt{vec}{\left[ {{\partial _{{\rm \kappa_{a}}}}\mathcal{C}} \right]^\dag }{\mathcal{Z}^{ - 1}}\mathtt{vec}\left[ {{\partial_{{\rm g}}}\mathcal{C} } \right] + {\partial _{{\rm \kappa_{a}}}}{\mathbf{r}^\top}\hspace{0.1cm}\mathcal{C}^{-1}\hspace{0.1cm}{\partial _{{\rm g}}}\mathbf{r} & 2\mathtt{vec}{\left[ {{\partial _{\kappa_{\rm a}}}\mathcal{C}} \right]^\dag }{\mathcal{Z}^{ - 1}}\mathtt{vec}\left[ {{\partial_{\kappa_{\rm a}}}\mathcal{C} } \right] + {\partial _{{\rm \kappa_{a}}}}{\mathbf{r}^\top}\hspace{0.1cm}\mathcal{C}^{-1}\hspace{0.1cm}{\partial _{{\rm \kappa_{a}}}}\mathbf{r}
\end{pmatrix}\nonumber\\
&&=\begin{pmatrix}
{\mathbb{H}}_{\rm gg} & {\mathbb{H}}_{\rm g\kappa_{\rm a}} \\
{\mathbb{H}}_{\kappa_{\rm a}{\rm g}} & {\mathbb{H}}_{\kappa_{\rm a}\kappa_{\rm a}}
\end{pmatrix}.\
\end{eqnarray}
Accordingly, $\mathcal{B}_{\rm S}$ and $\mathcal{B}_{\rm R}$ are obtained from Eqs.~\eqref{BS} and \eqref{SLDM}, and Eqs.~\eqref{RD} and \eqref{RLDM}, respectively.

\section{Results and discussions} \label{sec5}

In this section, we investigate the steady-state performance of multiparameter quantum estimation in the hybrid optomechanical system, with particular emphasis on the influence of experimentally accessible control parameters. Specifically, we examine their effects on the most informative quantum Cram\'er--Rao bound (QCRB), $\mathcal{B}_{\rm MI}$, which quantifies the ultimate precision limit for the simultaneous estimation of multiple parameters. We first analyze the role of the intracavity parametric amplifier (PA), focusing on its gain $\lambda$ and squeezing phase $\theta$, and subsequently investigate the effect of the coherent-feedback mechanism through the reflectivity parameter $\tau$. Finally, to assess the practical relevance of the quantum precision bounds, we consider experimentally realistic parameter regimes. Unless otherwise stated, all numerical results are obtained using the system parameters summarized in Table~\ref{table}.

\begin{table}[htbp]
\centering
\caption{Parameters of the optomechanical system used in the numerical calculations \cite{Groblasher09}.}
\label{table}
\begin{tabular*}{\textwidth}{@{\extracolsep{\fill}} l c c c}
\hline\hline
Parameter & Symbol & Value \\
\hline
Mechanical resonance frequency & $\omega_{\rm b}/2\pi$ & $947\times10^{3}\ {\rm Hz}$ \\
Mechanical damping rate & $\kappa_{\rm b}/2\pi$ & $140\ {\rm Hz}$ \\
Mirror mass & $\rm m$ & $145\ {\rm ng}$ \\
Cavity length & $\rm L$ & $25\ {\rm mm}$ \\
Cavity resonance frequency & $\omega_{\rm a}/2\pi$ & $5.26\times10^{14}\ {\rm Hz}$ \\
Laser frequency & $\omega_{\rm L}/2\pi$ & $2.82\times10^{14}\ {\rm Hz}$ \\
Laser wavelength & $\ell$ & $1064\ {\rm nm}$ \\
\hline\hline
\end{tabular*}
\end{table}

\subsection{Effect of parametric amplifier in the absence of CF}
\subsubsection{\bf{Steady-state}}

\begin{figure}[!htb]
\includegraphics[width=0.45\linewidth]{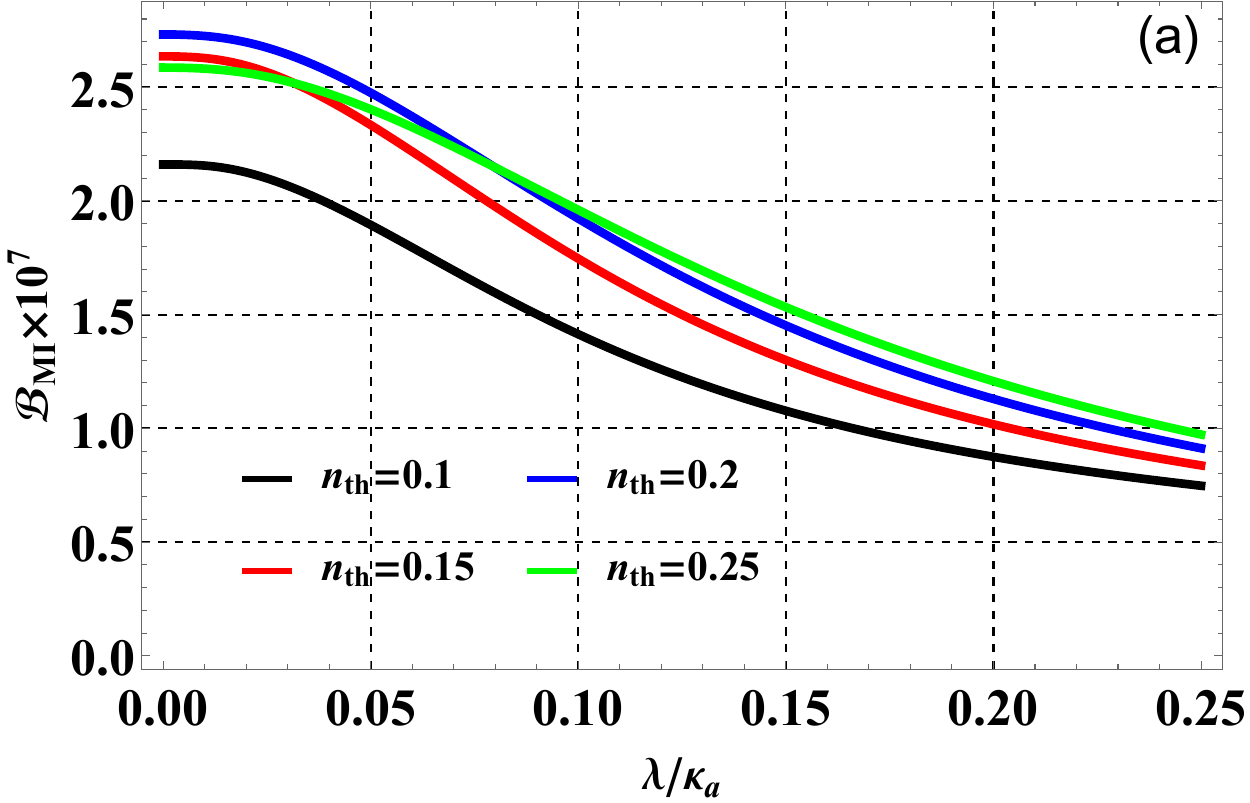}
\includegraphics[width=0.45\linewidth]{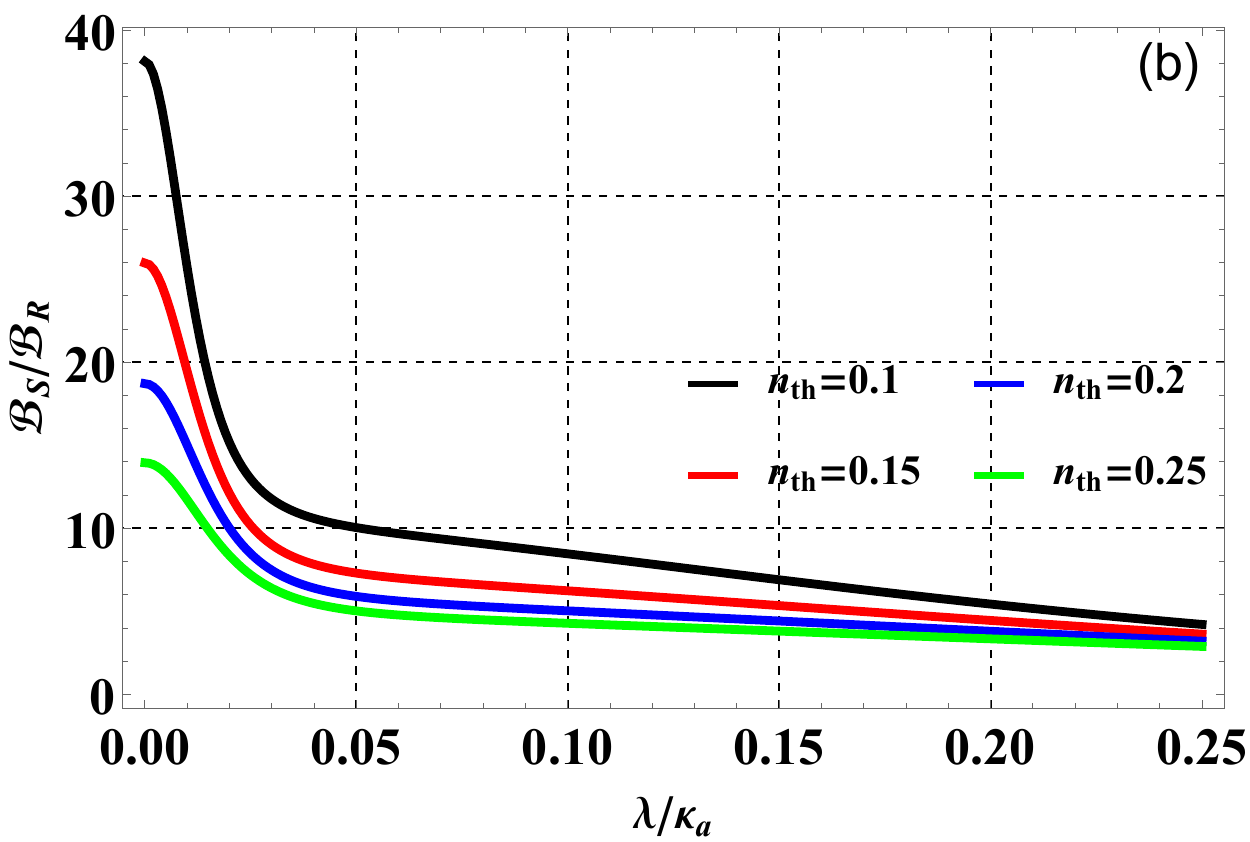}
\caption{Plots of (a) the most informative quantum Cram\'er--Rao bound, $\mathcal{B}_{\rm MI}$, and (b) the ratio of the SLD- to RLD-based bounds, $\mathcal{B}_{\rm S}/\mathcal{B}_{\rm R}$, as functions of the normalized parametric gain $\lambda/\kappa_{\rm a}$ for different values of the mean thermal photon number $n_{\rm th}$. The parametric-amplifier phase and the squeezing parameter are fixed at $\theta=\pi/2$ and $\rm r=0.1$, respectively, while all other system parameters are chosen according to the values listed in Table~\ref{table}.}
\label{fig2}
\end{figure}
In Fig.~\ref{fig2}, we investigate the influence of the parametric-amplifier (PA) gain on the precision of simultaneous multiparameter quantum estimation. Figure~\ref{fig2}(a) presents the most informative quantum Cram\'er--Rao bound (QCRB), $\mathcal{B}_{\rm MI}$, whereas Fig.~\ref{fig2}(b) shows the ratio of the SLD- and RLD-based bounds $\mathcal{B}_{\rm S}/\mathcal{B}_{\rm R}$, as functions of the normalized PA gain $\lambda/\kappa_{\rm a}$ for different values of the mean thermal photon number $n_{\rm th}$. The quantity $\mathcal{B}_{\rm MI}$ characterizes the tightest precision bound considered here, while the ratio $\mathcal{B}_{\rm S}/\mathcal{B}_{\rm R}$ provides complementary information about the relative behavior of the SLD- and RLD-based bounds.

As shown in Fig.~\ref{fig2}(a), $\mathcal{B}_{\rm MI}$ decreases systematically with increasing $\lambda/\kappa_{\rm a}$ for all considered values of $n_{\rm th}$. Since a smaller QCRB corresponds to a higher attainable estimation precision, this behavior demonstrates that increasing the PA gain enhances the performance of multiparameter estimation. The physical origin of this enhancement lies in the modification of the quantum fluctuations and intermode correlations induced by parametric amplification. By appropriately reshaping the covariance matrix, the PA can strengthen the parameter-dependent correlations that encode information about the unknown quantities. This increases the distinguishability of quantum states associated with different parameter values and consequently enhances the quantum information available for estimation.

The gain dependence, however, is not uniform over the entire parameter range. The reduction of $\mathcal{B}_{\rm MI}$ is more pronounced at low-to-intermediate values of $\lambda/\kappa_{\rm a}$, while the improvement becomes progressively weaker as the gain is further increased. This saturation-like behavior indicates that the information enhancement provided by the PA is subject to diminishing returns. At moderate gain, the amplification efficiently enhances the correlations relevant to parameter estimation; at larger gain, the simultaneous amplification of quantum fluctuations increasingly offsets the additional benefit obtained from the enhanced correlations. Therefore, increasing the PA gain beyond a certain range does not lead to a proportional improvement in the estimation precision.

The influence of thermal noise is also clearly visible in Fig.~\ref{fig2}(a). For a given value of $\lambda/\kappa_{\rm a}$, increasing $n_{\rm th}$ leads to larger values of $\mathcal{B}_{\rm MI}$, corresponding to a degradation of the estimation precision. Thermal excitation introduces additional fluctuations and increases the mixedness of the Gaussian state, thereby weakening the parameter-dependent quantum correlations and reducing the distinguishability between neighboring states in parameter space. As a result, the amount of useful quantum information available for estimating the unknown parameters is reduced. Nevertheless, $\mathcal{B}_{\rm MI}$ continues to decrease with increasing PA gain even for finite thermal occupation, showing that parametric amplification can partially compensate for the detrimental effect of thermal noise.

Figure~\ref{fig2}(b) complements this analysis by examining the ratio $\mathcal{B}_{\rm S}/\mathcal{B}_{\rm R}$. At small PA gain, the ratio changes rapidly with $\lambda/\kappa_{\rm a}$, indicating that parametric amplification strongly modifies the relative behavior of the SLD and RLD-based precision bounds. This effect originates from the gain-induced modification of the covariance matrix and, consequently, of the quantum statistical structure governing the simultaneous estimation of the parameters. Thus, the role of the PA is not restricted to increasing the overall amount of quantum information; it also changes the relative tightness of the two quantum Cram\'er--Rao bounds.

For $\lambda\gtrsim0.05\kappa_{\rm a}$, the variation of $\mathcal{B}_{\rm S}/\mathcal{B}_{\rm R}$ becomes much weaker, indicating that the dominant gain-induced modification of the relative SLD and RLD bounds occurs in the low-to-intermediate-gain regime. Further increasing the gain produces only a minor change in their relative behavior. This trend is consistent with the saturation observed in $\mathcal{B}_{\rm MI}$ and suggests that the principal metrological advantage of the PA is obtained before entering the high-gain regime.

Thermal fluctuations also influence the ratio $\mathcal{B}_{\rm S}/\mathcal{B}_{\rm R}$. In particular, increasing $n_{\rm th}$ leads to a lower ratio over the investigated gain range, demonstrating that thermal noise modifies not only the absolute estimation precision but also the relative performance of the SLD- and RLD-based bounds. This behavior reflects the modification of the covariance and mixedness of the Gaussian state by thermal excitation, which consequently changes the quantum statistical geometry relevant to multiparameter estimation.

Overall, Fig.~\ref{fig2} reveals a clear interplay between parametric amplification and thermal noise. Increasing the PA gain improves the attainable estimation precision, as evidenced by the reduction of $\mathcal{B}_{\rm MI}$, while simultaneously modifying the relative behavior of the SLD and RLD bounds. In contrast, thermal fluctuations degrade the overall precision and alter the relative structure of the two bounds. Importantly, the strongest metrological enhancement is obtained in the low-to-intermediate-gain regime, whereas the benefit becomes progressively weaker at larger gain. This identifies the PA gain as an effective control parameter for optimizing multiparameter quantum sensing, with an appropriate gain providing a favorable balance between the enhancement of useful quantum correlations and the amplification of unwanted quantum fluctuations.\\

\begin{figure}[!htb]
\includegraphics[width=0.5\linewidth]{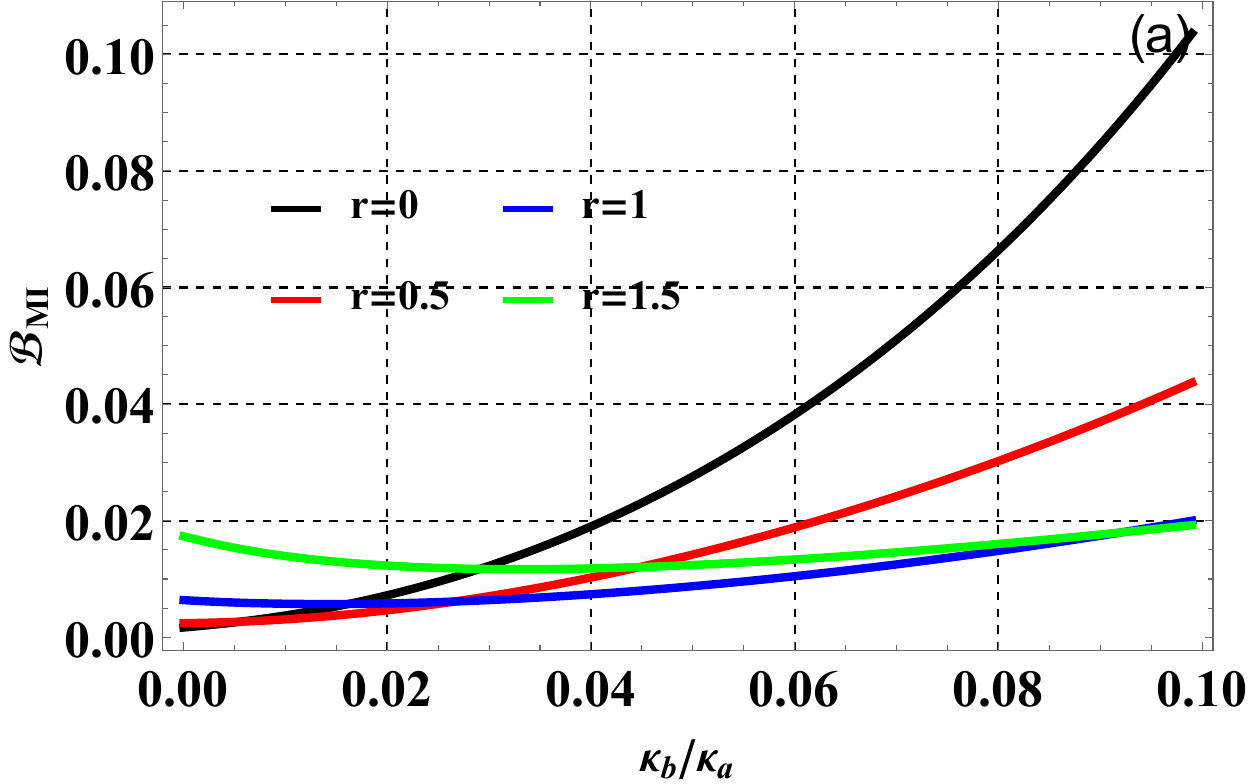}
\includegraphics[width=0.47\linewidth]{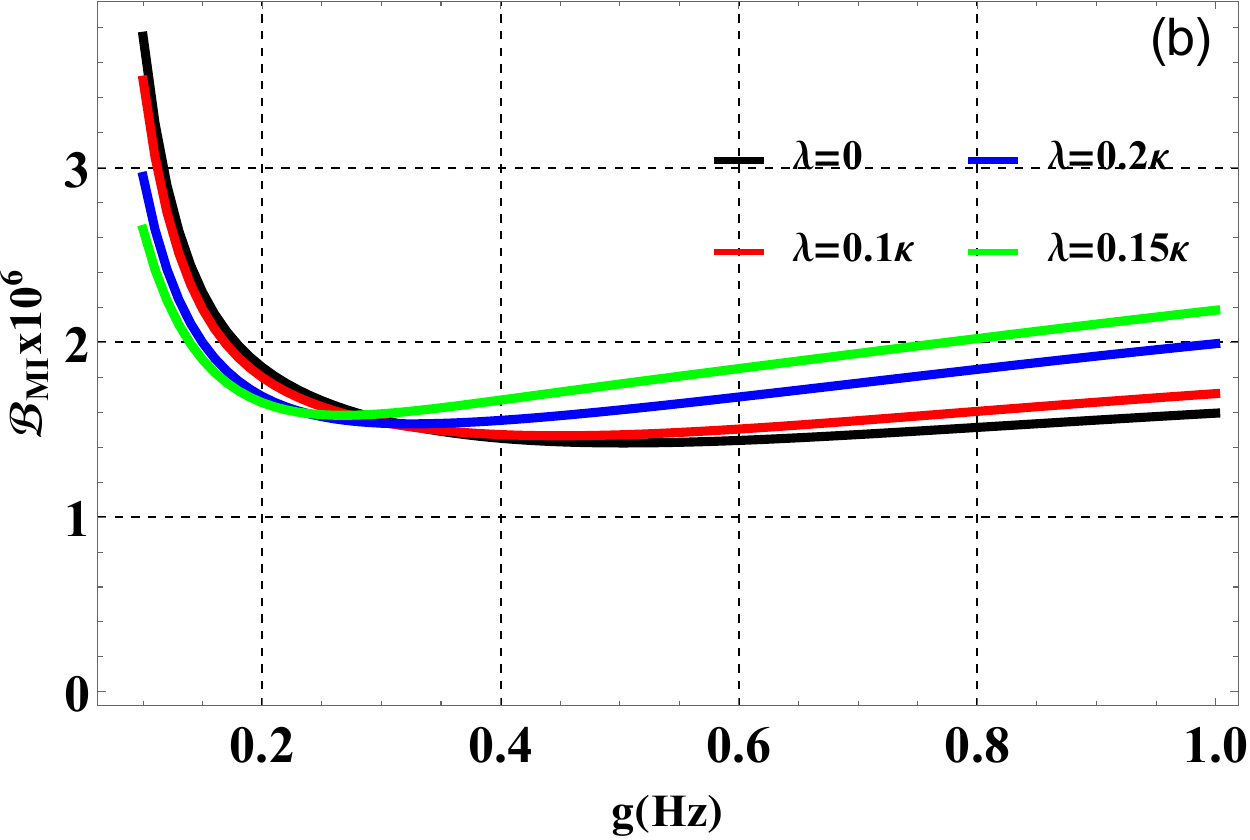}
\caption{Plots of the most informative quantum Cram\'er--Rao bound (QCRB), $\mathcal{B}_{\rm MI}$, as a function of (a) the normalized mechanical damping rate $\kappa_{\rm b}/\kappa_{\rm a}$ for different values of the squeezing parameter $\rm r$, with $\lambda=0.25\kappa_{\rm a}$, and (b) the optomechanical coupling strength $\rm g$ (Hz) for different values of the parametric-amplifier gain $\lambda$, with $\rm r=0.1$. In both panels, the parametric-amplifier phase is fixed at $\theta=\pi/2$ and the mean thermal photon number is set to $n_{\rm th}=0.1$. All other system parameters are chosen according to the values listed in Table~\ref{table}.}
\label{fig3}
\end{figure}
In Fig.~\ref{fig3}, we investigate the dependence of the most informative quantum Cram\'er--Rao bound (QCRB), $\mathcal{B}_{\rm MI}$, on the damping rate and coupling strength under different squeezing and PA-gain conditions. Figure~\ref{fig3}(a) shows $\mathcal{B}_{\rm MI}$ as a function of the normalized damping rate $\kappa_{\rm b}/\kappa_{\rm a}$ for different values of the squeezing parameter $\rm r$, while Fig.~\ref{fig3}(b) presents its variation with the coupling strength $\rm g$ for different values of the PA gain $\lambda$.
As shown in Fig.~\ref{fig3}(a), $\mathcal{B}_{\rm MI}$ increases with $\kappa_{\rm b}/\kappa_{\rm a}$, indicating that stronger damping deteriorates the multiparameter estimation precision. This degradation results from the enhanced dissipation and associated fluctuations, which reduce the useful quantum information available for parameter estimation. Interestingly, the effect of squeezing depends on the damping regime. For $\kappa_{\rm b}<0.2\kappa_{\rm a}$, increasing $\rm r$ leads to an increase in $\mathcal{B}_{\rm MI}$, whereas for $\kappa_{\rm b}>0.2\kappa_{\rm a}$, the opposite behavior is observed, with $\mathcal{B}_{\rm MI}$ decreasing as $\rm r$ increases. This crossover indicates a competition between squeezing-induced modifications of the quantum fluctuations and dissipative effects. Thus, squeezing becomes beneficial for estimation in the stronger-damping regime, where it can partially compensate for the information loss induced by dissipation.
In Fig.~\ref{fig3}(b), $\mathcal{B}_{\rm MI}$ decreases with increasing coupling strength for $\rm g<0.3$~Hz, demonstrating that stronger coupling enhances the transfer of parameter-dependent quantum correlations and consequently improves the estimation precision. In this weak-coupling regime, increasing the PA gain $\lambda$ further reduces $\mathcal{B}_{\rm MI}$, showing that parametric amplification provides an additional enhancement of the available quantum information. However, for $\rm g\gtrsim0.3$--$0.4$~Hz, $\mathcal{B}_{\rm MI}$ becomes nearly insensitive to further increases in $\rm g$, indicating that the information transfer has essentially reached a saturation regime.
Moreover, the effect of the PA gain changes with the coupling strength. For $\rm g<0.3$~Hz, increasing $\lambda$ reduces $\mathcal{B}_{\rm MI}$, whereas for stronger coupling, $\mathcal{B}_{\rm MI}$ increases with $\lambda$. This reversal suggests a competition between the beneficial gain-induced enhancement of quantum correlations and the additional fluctuations introduced by parametric amplification. Consequently, the PA is most advantageous in the weak-coupling regime, while excessive gain can become detrimental when the coupling is already sufficiently strong.
Overall, Fig.~\ref{fig3} demonstrates that the estimation precision is governed by a nontrivial interplay between dissipation, squeezing, coupling, and parametric amplification. In particular, the results indicate that the optimal estimation performance is achieved by appropriately balancing these parameters rather than simply maximizing the squeezing, coupling strength, or PA gain individually.\\

\begin{figure}[!htb]
\includegraphics[width=0.45\linewidth]{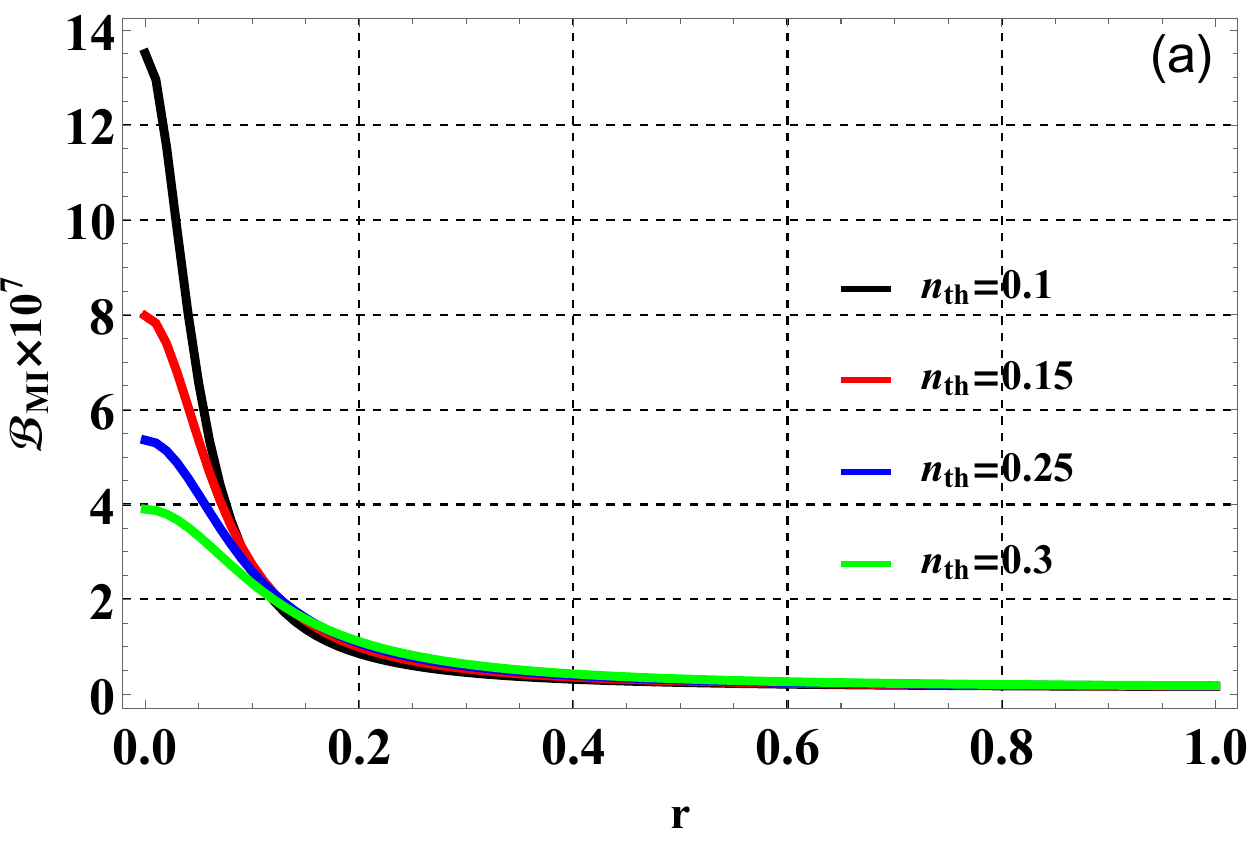}
\includegraphics[width=0.45\linewidth]{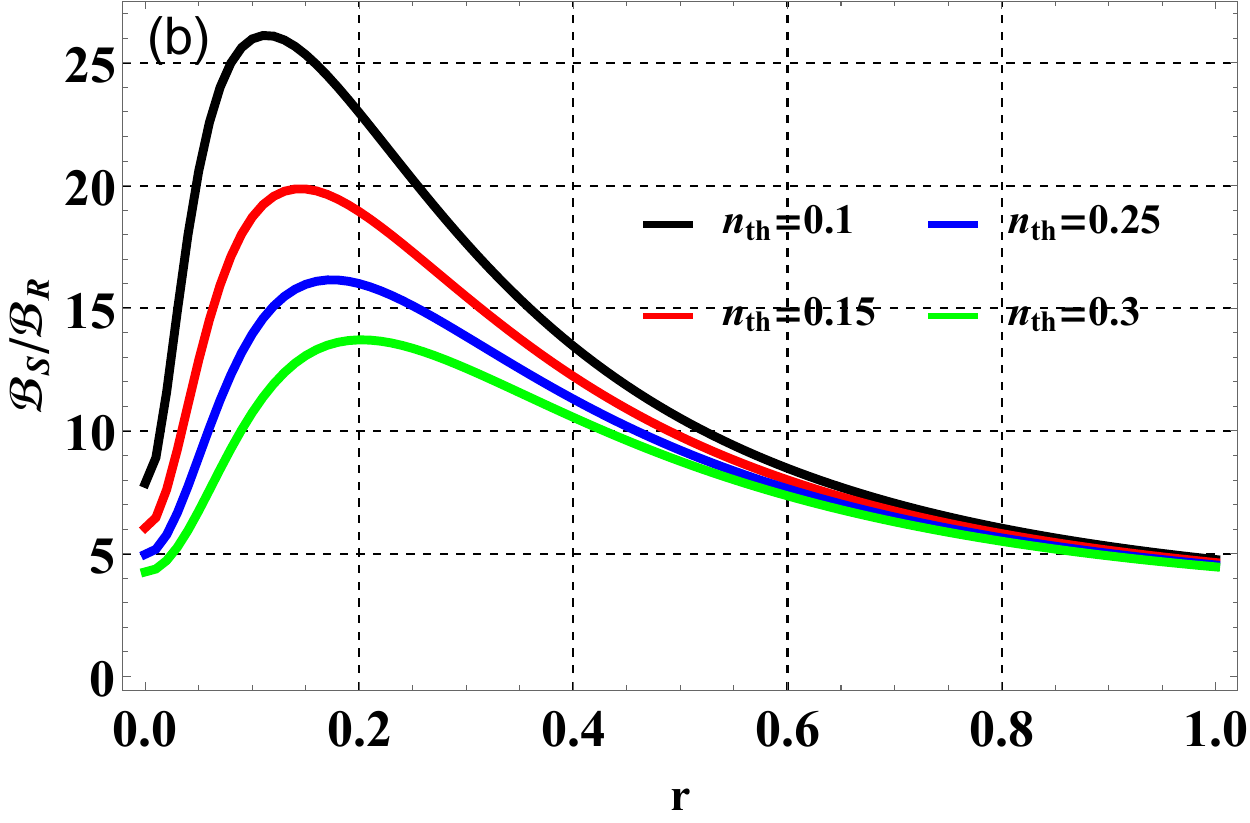}
\caption{Plot of the: (a) most informative quantum Cram\'er--Rao bound (QCRB) $\mathcal{B}_{\rm MI}$ and (b) the ratio of the SLD- and RLD-based quantum Cram\'er--Rao bounds, $\mathcal{B}_{\rm S}/\mathcal{B}_{\rm R}$ as function of the squeezing parameter $\rm r$  for different values of the $n_{\rm th}$, with $\lambda=0.25\kappa_{\rm a}$ and $\theta=\frac{\pi}{2}$. The other parameters are provided in the table \ref{table}.}
\label{fig5}
\end{figure}
In Fig.~\ref{fig5}, we investigate the influence of the squeezing parameter $\rm r$ on the multiparameter estimation precision through the most informative quantum Cram\'er--Rao bound (QCRB), $\mathcal{B}_{\rm MI}$, in panel (a), and the ratio of the SLD- and RLD-based bounds, $\mathcal{B}_{\rm S}/\mathcal{B}_{\rm R}$, in panel (b), for different values of the mean thermal photon number $n_{\rm th}$. As shown in Fig.~\ref{fig5}(a), $\mathcal{B}_{\rm MI}$ decreases for small values of $\rm r$, indicating that increasing the squeezing strength enhances the estimation precision by increasing the parameter-dependent quantum correlations and, consequently, the accessible information. As $n_{\rm th}$ increases, $\mathcal{B}_{\rm MI}$ decreases, reflecting the detrimental effect of thermal fluctuations, which reduce the distinguishability of the parameter-dependent quantum states. For ${\rm r}>0.6$, however, $\mathcal{B}_{\rm MI}$ becomes nearly insensitive to variations in $n_{\rm th}$, suggesting that the squeezing-induced correlations become dominant over the influence of thermal noise in this regime.

Figure~\ref{fig5}(b) shows a different behavior for $\mathcal{B}_{\rm S}/\mathcal{B}_{\rm R}$. For weak squeezing, the ratio initially increases with $\rm r$, while for $\rm r>0.1$ it decreases rapidly, indicating that squeezing significantly modifies the relative behavior of the SLD and RLD-based bounds. This variation reflects the modification of the quantum statistical structure and parameter correlations induced by squeezing. Moreover, increasing $n_{\rm th}$ systematically reduces $\mathcal{B}_{\rm S}/\mathcal{B}_{\rm R}$, showing that thermal fluctuations alter not only the amount of available information but also the relative tightness of the two quantum estimation bounds. Overall, these results demonstrate that squeezing can substantially enhance multiparameter estimation, while its interplay with thermal noise determines the ultimate estimation performance.  

\begin{figure}[!htb]
\includegraphics[width=0.45\linewidth]{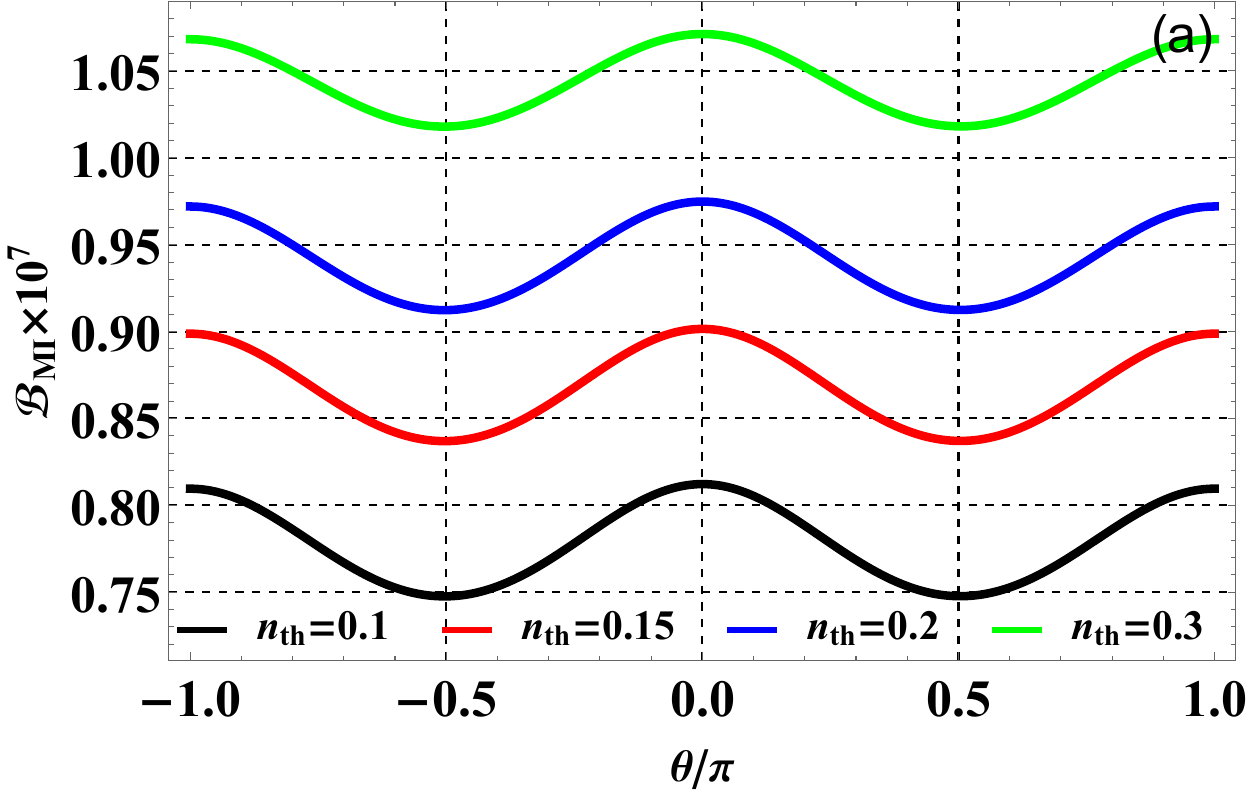}
\includegraphics[width=0.45\linewidth]{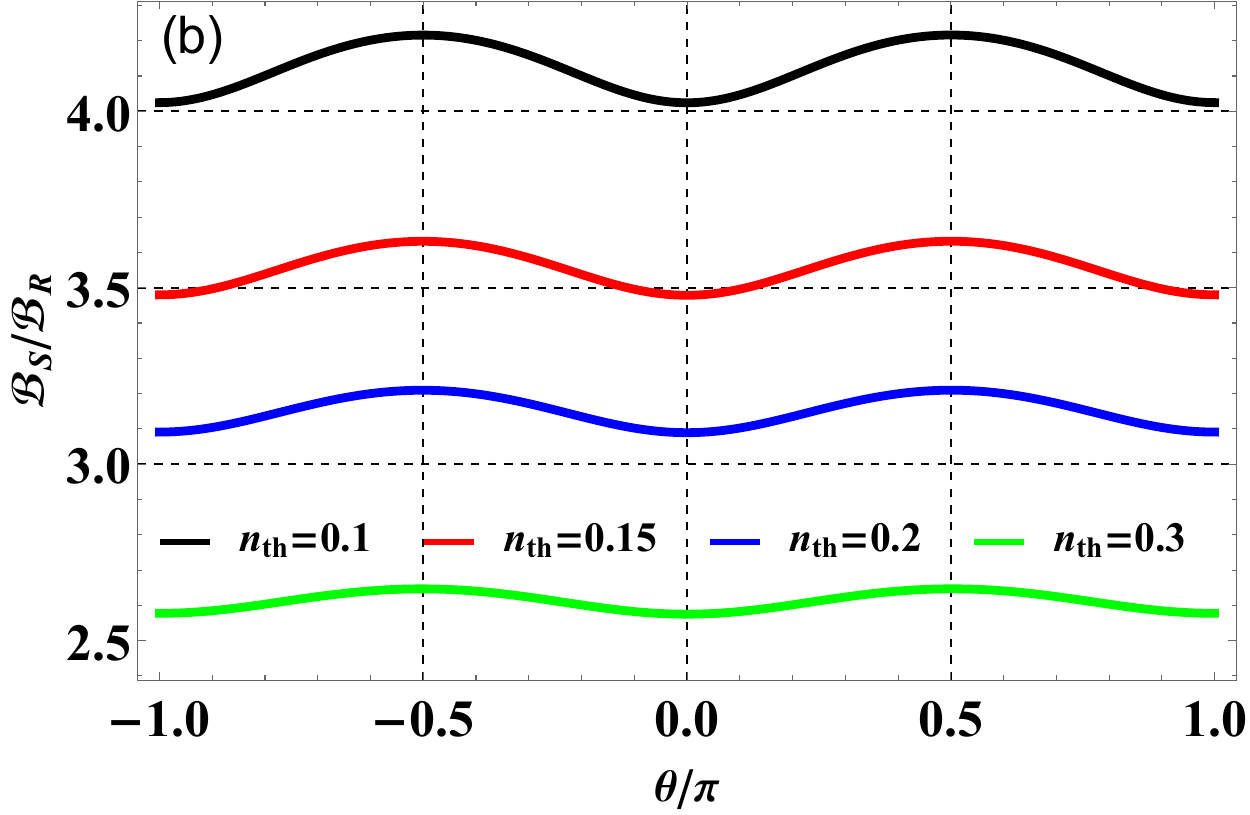}
\caption{Plot of the: (a) most informative quantum Cram\'e
r--Rao bound (QCRB) $\mathcal{B}_{\rm MI}$ and (b) ratio between the symmetric logarithmic derivative (SLD) and right logarithmic derivative (RLD) bounds, $\mathcal{B}_{\rm S}/\mathcal{B}_{\rm R}$ as function of the normalized phase $\theta/\pi$ for different values of the mean thermal photon number $n_{\rm th}$, with $\lambda=0.25\kappa_{\rm a}$ and the squeeze parameter is fixed at $\rm r=0.1$. The other parameters are provided in the table \ref{table}.}
\label{fig6}
\end{figure}
In Fig.~\ref{fig6}, we investigate the influence of the PA phase $\theta$ on the multiparameter estimation performance by presenting the most informative quantum Cram\'er--Rao bound (QCRB), $\mathcal{B}_{\rm MI}$, in panel (a), and the ratio of the SLD and RLD-based bounds $\mathcal{B}_{\rm S}/\mathcal{B}_{\rm R}$, in panel (b), as functions of the normalized phase $\theta/\pi$ for different values of the mean thermal photon number $n_{\rm th}$. As shown in Fig.~\ref{fig6}(a), $\mathcal{B}_{\rm MI}$ exhibits a pronounced periodic dependence on the PA phase $\theta$ for all considered thermal occupations. This periodic behavior originates from the phase-sensitive nature of the parametric amplification process, which modifies the quadrature fluctuations and quantum correlations depending on the relative phase of the PA. Consequently, the amount of information encoded in the quantum state and, hence, the estimation precision can be periodically enhanced or degraded by tuning $\theta$. In addition, increasing $n_{\rm th}$ leads to larger values of $\mathcal{B}_{\rm MI}$, indicating that thermal fluctuations reduce the available quantum information and deteriorate the multiparameter estimation precision. Nevertheless, the periodic phase dependence remains visible, showing that the PA phase can be used as an effective control parameter for optimizing the estimation performance even in the presence of thermal noise.

Figure~\ref{fig6}(b) shows that the ratio $\mathcal{B}_{\rm S}/\mathcal{B}_{\rm R}$ also exhibits a clear periodic dependence on $\theta/\pi$. This behavior indicates that the PA phase not only affects the overall estimation precision but also modifies the relative behavior of the SLD- and RLD-based quantum bounds. Such phase sensitivity results from the modification of the quantum statistical structure and parameter correlations induced by the phase-dependent parametric amplification. Moreover, increasing the thermal photon number $n_{\rm th}$ systematically reduces $\mathcal{B}_{\rm S}/\mathcal{B}_{\rm R}$, demonstrating that thermal fluctuations influence the relative tightness of the two quantum estimation bounds. Overall, these results highlight the PA phase as an additional control parameter that can be exploited to tune both the amount of accessible quantum information and the structure of the multiparameter estimation bounds.

\subsubsection{\bf{Dynamical state}}

\begin{figure}[!htb]
\includegraphics[width=0.45\linewidth]{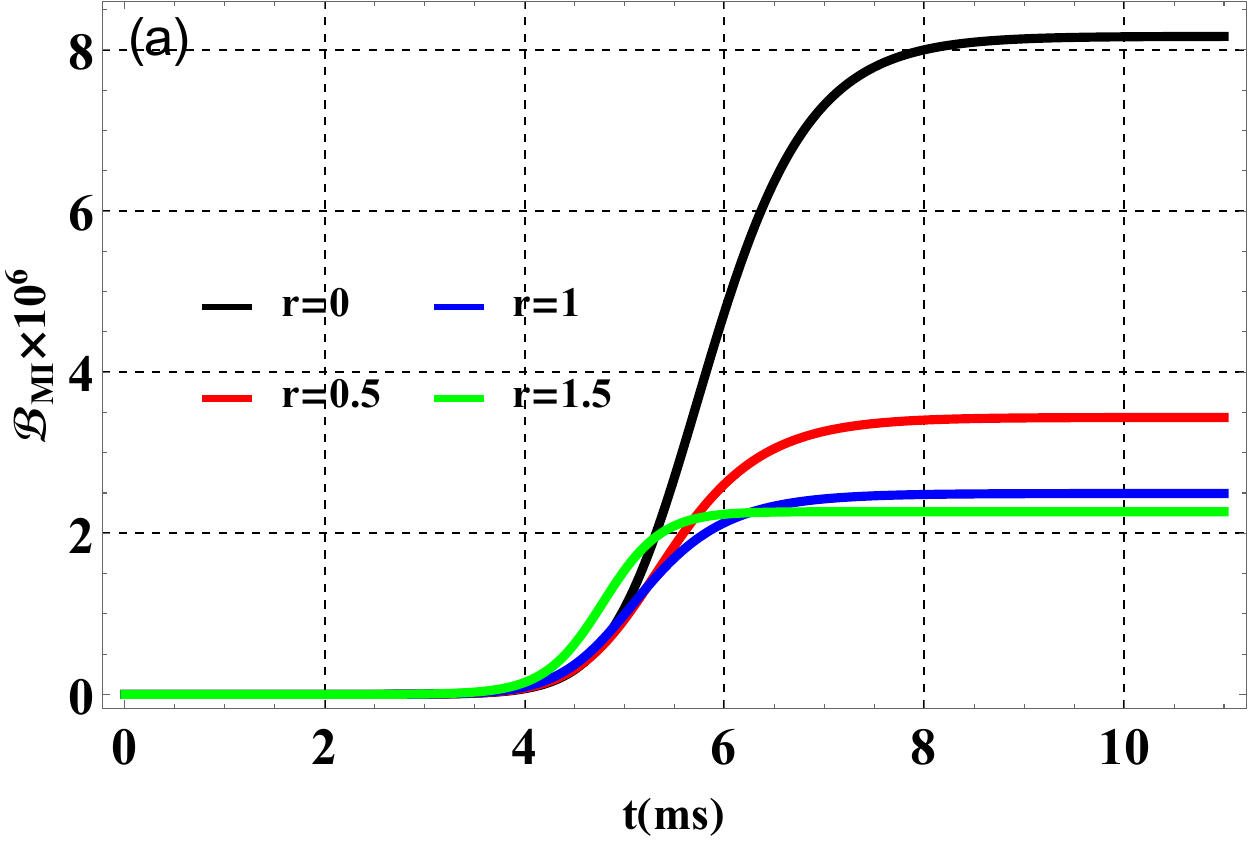}
\includegraphics[width=0.45\linewidth]{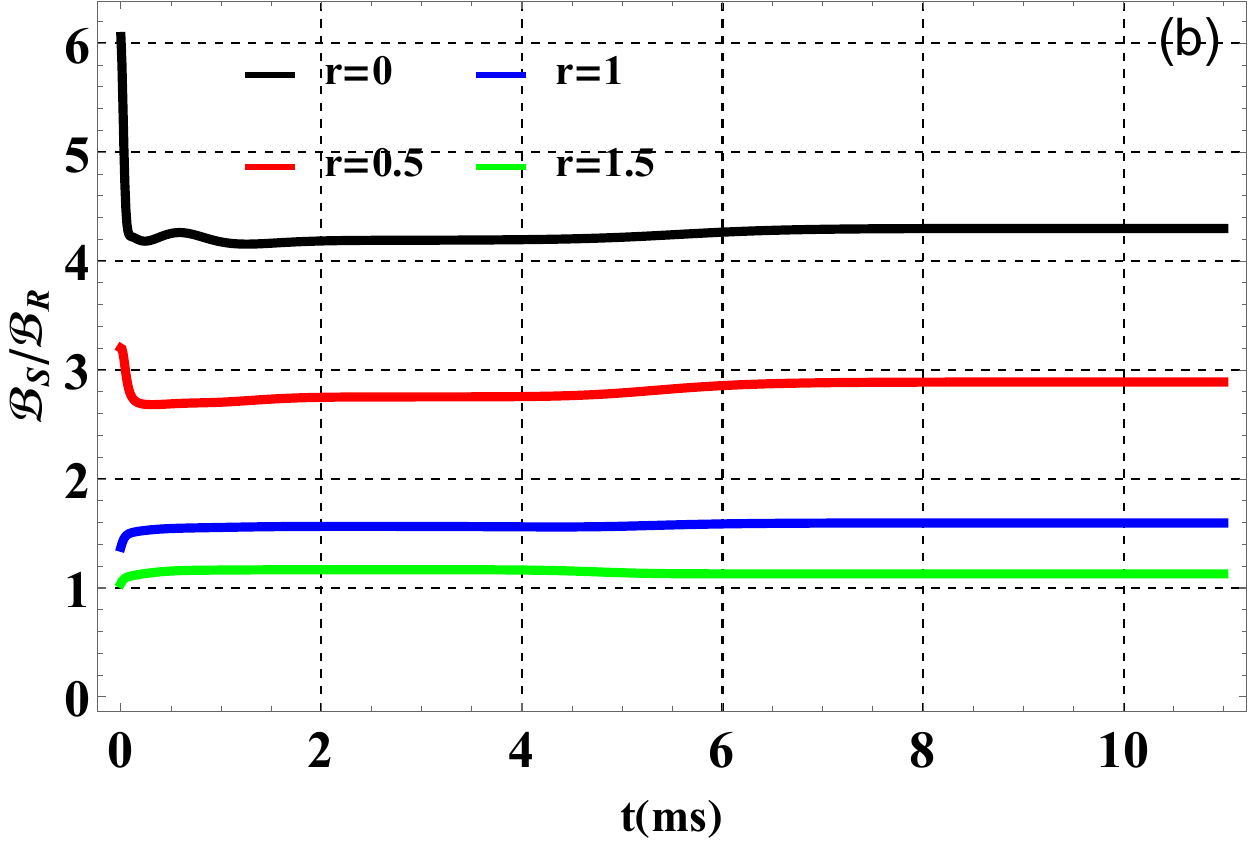}
\caption{Plot of: the evolution of (a) the most informative quantum Cram\'er--Rao bound (QCRB) $\mathcal{B}_{\rm MI}$ and (b) the ratio of the SLD- and RLD-based bounds, $\mathcal{B}_{\rm S}/\mathcal{B}_{\rm R}$  for different values of the squeeze parameter $\rm r$, with the gain of PA $\lambda=0.25\kappa_{\rm a}$, the phase is fixed at $\theta=\frac{\pi}{2}$ and the mean thermal phonton is fixed at $n_{\rm th}=0.1$. The other parameters are provided in the table \ref{table}.}
\label{fig7}
\end{figure}

Figure~\ref{fig7} illustrates the temporal evolution of the most informative quantum Cram\'er--Rao bound (QCRB), $\mathcal{B}_{\rm MI}$, together with the ratio between the symmetric logarithmic derivative (SLD) and right logarithmic derivative (RLD) bounds, $\mathcal{B}_{\rm S}/\mathcal{B}_{\rm R}$, for different values of the squeezing parameter $\rm r$. As shown in Fig.~\ref{fig7}(a), the most informative QCRB remains essentially zero for short evolution times, up to approximately $\rm t\simeq4\,\mathrm{ms}$. This behavior indicates that, during the initial stage of the dynamics, the system does not yet provide appreciable information about the parameters to be estimated. For $\rm t>4\,\mathrm{ms}$, $\mathcal{B}_{\rm MI}$ progressively increases with time, indicating a gradual degradation of the estimation precision, since the estimation uncertainty is directly related to the QCRB. At longer times, namely for $\rm t\gtrsim8\,\mathrm{ms}$, $\mathcal{B}_{\rm MI}$ approaches an almost stationary regime, suggesting that the system reaches a quasi-steady estimation performance in which further evolution produces only a negligible change in the attainable precision.

Moreover, Fig.~\ref{fig7}(a) clearly shows that increasing the squeezing parameter $\rm r$ leads to a reduction of $\mathcal{B}_{\rm MI}$. Since a smaller QCRB corresponds to a lower fundamental estimation uncertainty, this result demonstrates that stronger squeezing enhances the precision of multiparameter estimation. Physically, the injected squeezed field modifies the quantum fluctuations and correlations of the system, allowing more useful information about the encoded parameters to be extracted from the output state. Thus, squeezing acts as a resource for improving the sensitivity of the quantum estimation protocol.

In Fig.~\ref{fig7}(b), we investigate the ratio $\mathcal{B}_{\rm S}/\mathcal{B}_{\rm R}$, which provides a useful measure for comparing the SLD- and RLD-based precision bounds. We observe that this ratio remains approximately constant over most of the evolution time, indicating that the relative behavior of the two bounds is largely insensitive to the dynamical evolution once the transient regime has passed. At very short times, however, a noticeable decrease of $\mathcal{B}_{\rm S}/\mathcal{B}_{\rm R}$ is observed for the unsqueezed case, $\rm r=0$. Furthermore, increasing the squeezing parameter leads to a reduction of the ratio $\mathcal{B}_{\rm S}/\mathcal{B}_{\rm R}$. This behavior indicates that squeezing modifies the relative tightness of the SLD and RLD bounds and, consequently, affects the degree of compatibility between the precision limits associated with the two logarithmic-derivative approaches. In particular, the reduction of $\mathcal{B}_{\rm S}/\mathcal{B}_{\rm R}$ with increasing $\rm r$ suggests that the RLD bound becomes relatively more significant compared with the SLD bound in the presence of stronger squeezing.

Overall, these results demonstrate that the squeezing parameter plays a dual role in the multiparameter estimation process: it significantly reduces the most informative QCRB, thereby enhancing the attainable estimation precision, while simultaneously modifying the relative relationship between the SLD and RLD bounds. The combined behavior highlights the importance of quantum squeezing not only as a resource for increasing the amount of information encoded in the system, but also for shaping the fundamental precision limits and measurement compatibility of multiparameter quantum estimation.

\begin{figure}[!htb]
\includegraphics[width=0.45\linewidth]{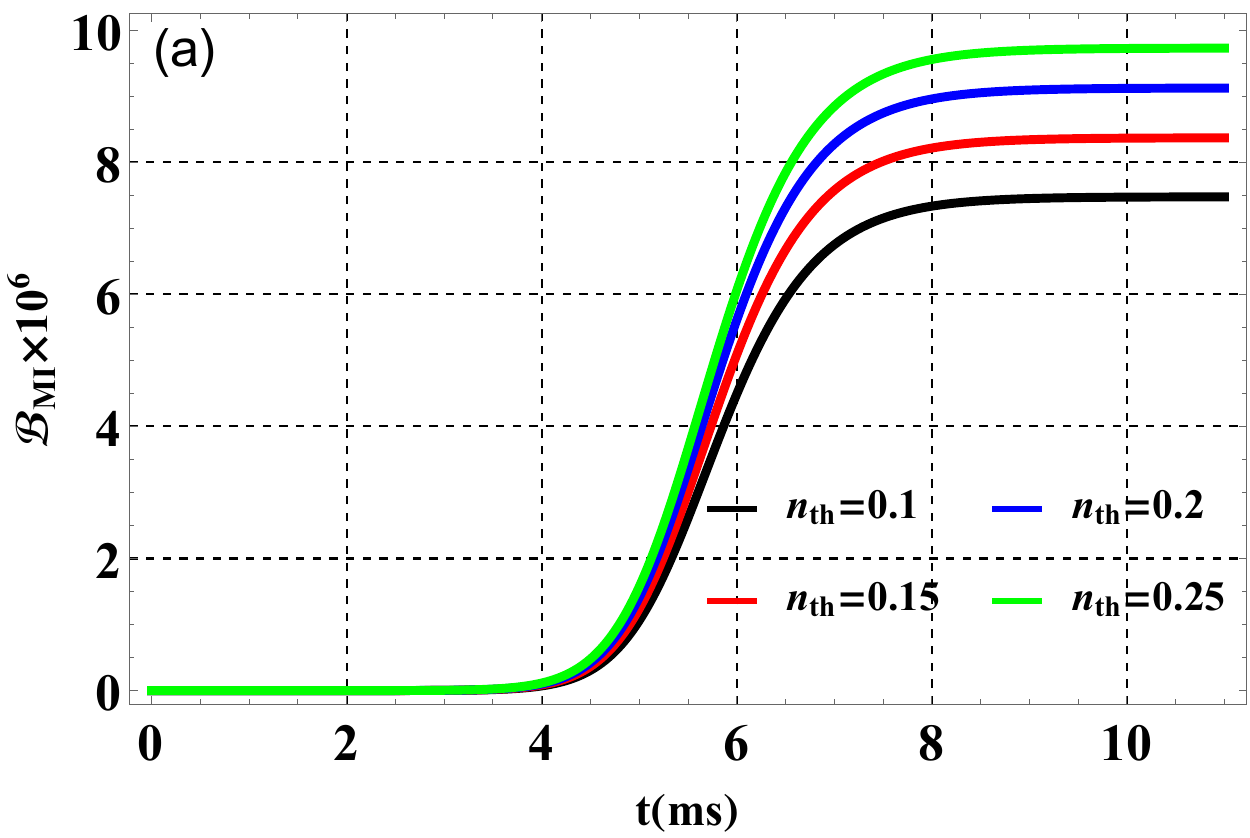}
\includegraphics[width=0.45\linewidth]{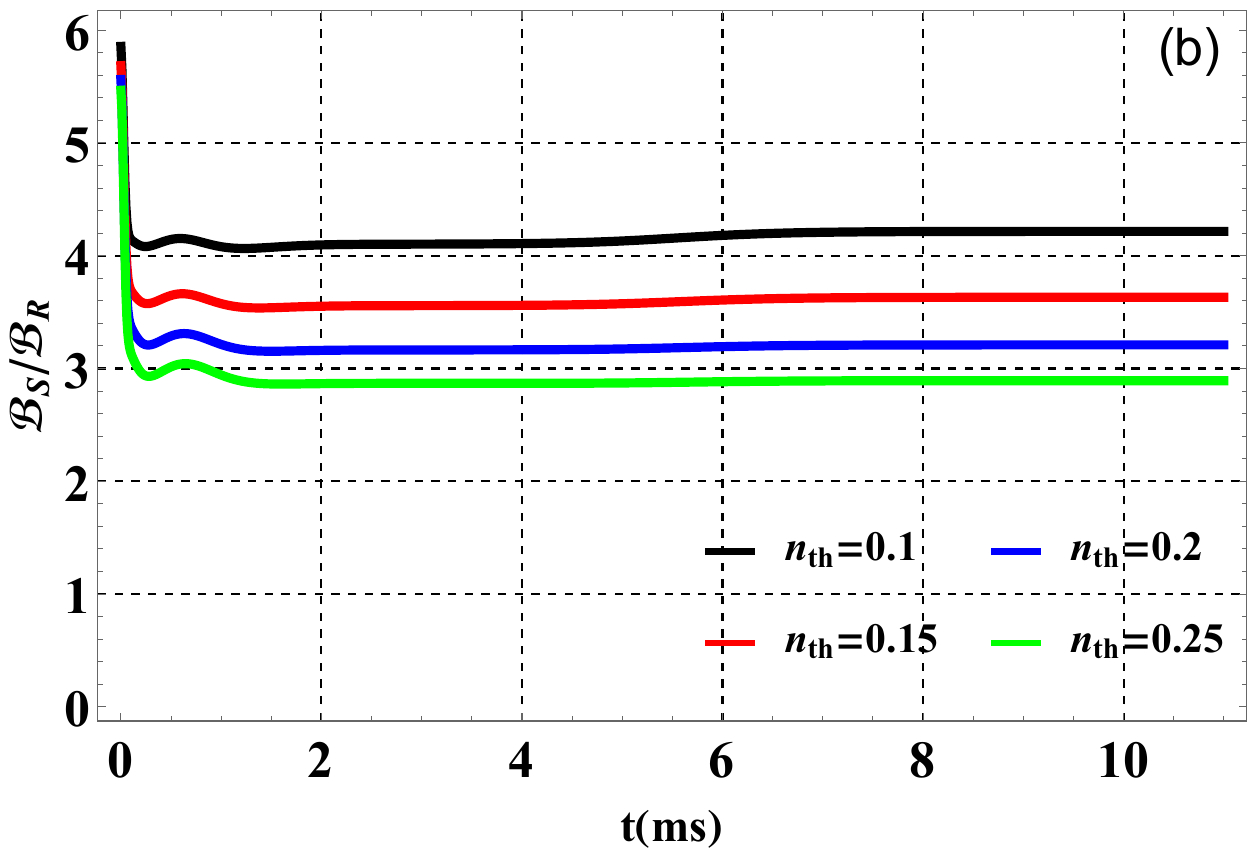}
\caption{Plot of: the evolution of (a) the most informative quantum Cram\'er--Rao bound (QCRB) $\mathcal{B}_{\rm MI}$ and (b)  the ratio between the symmetric logarithmic derivative (SLD) and right logarithmic derivative (RLD) bounds, $\mathcal{B}_{\rm S}/\mathcal{B}_{\rm R}$  for different values of the mean thermal photon number $n_{\rm th}$, with the gain of PA is fixed at $\lambda=0.25\kappa_{\rm a}$ and the phase is fixed at $\theta=\frac{\pi}{2}$ and the squeeze parameter is $\rm r=0.1$. The other parameters are provided in the table \ref{table}.}
\label{fig8}
\end{figure}
Figure~\ref{fig8} presents the dynamical evolution of (a) the most informative quantum Cram\'er--Rao bound (QCRB), $\mathcal{B}_{\rm MI}$, and (b) the ratio between the symmetric logarithmic derivative (SLD) and right logarithmic derivative (RLD) bounds, $\mathcal{B}_{\rm S}/\mathcal{B}_{\rm R}$, for different values of the mean thermal photon number $n_{\rm th}$. In Fig.~\ref{fig8}(a), we observe that $\mathcal{B}_{\rm MI}$ remains negligible during the initial stage of the dynamics and becomes appreciable for evolution times $t\gtrsim4\,\mathrm{ms}$. This behavior indicates that the system initially provides only a limited amount of useful information about the parameters to be estimated, whereas the estimation uncertainty becomes more pronounced at later times. As the evolution proceeds, $\mathcal{B}_{\rm MI}$ increases, corresponding to a degradation of the attainable estimation precision. Moreover, increasing the mean thermal photon number $n_{\rm th}$ leads to a systematic enhancement of $\mathcal{B}_{\rm MI}$. Since the QCRB sets a lower bound on the estimation variance, this increase indicates a deterioration of the parameter-estimation precision in the presence of stronger thermal noise. Physically, thermal excitations introduce additional fluctuations into the mechanical degrees of freedom, which contaminate the quantum correlations carrying information about the unknown parameters. Consequently, increasing $n_{\rm th}$ reduces the amount of useful information that can be extracted from the system and therefore weakens its metrological performance.

For Fig.~\ref{fig8}(b), the ratio $\mathcal{B}_{\rm S}/\mathcal{B}_{\rm R}$ exhibits a transient behavior at short evolution times, characterized by a decrease as the system evolves from its initial state. After this initial transient regime, the ratio approaches an approximately constant value, indicating that the relative behavior of the SLD and RLD bounds becomes nearly time-independent at longer evolution times. Furthermore, we observe that increasing the mean thermal photon number $n_{\rm th}$ leads to a reduction of $\mathcal{B}_{\rm S}/\mathcal{B}_{\rm R}$. This result demonstrates that thermal noise not only degrades the overall estimation precision, as reflected by the increase of $\mathcal{B}_{\rm MI}$, but also modifies the relative tightness of the SLD and RLD precision bounds. In particular, the reduction of $\mathcal{B}_{\rm S}/\mathcal{B}_{\rm R}$ with increasing $n_{\rm th}$ indicates that thermal fluctuations significantly affect the balance between the two quantum Fisher-information-based bounds and may enhance the relevance of the RLD bound in characterizing the ultimate precision limits of the multiparameter estimation protocol.

Overall, Fig.~\ref{fig8} demonstrates that thermal noise has a detrimental impact on the metrological performance of the system. An increase in $n_{\rm th}$ simultaneously increases the most informative QCRB and reduces the ratio $\mathcal{B}_{\rm S}/\mathcal{B}_{\rm R}$, revealing that thermal excitations degrade the attainable estimation precision and alter the relative behavior of the SLD and RLD bounds. These results emphasize the importance of operating the system at sufficiently low thermal occupation in order to preserve the quantum resources and correlations responsible for enhanced multiparameter estimation.

\subsection{Effect of the coherent feedback}

\begin{figure}[!htb]
\centering 
\begin{tabular}{ccc} 
\includegraphics[width=0.32\linewidth]{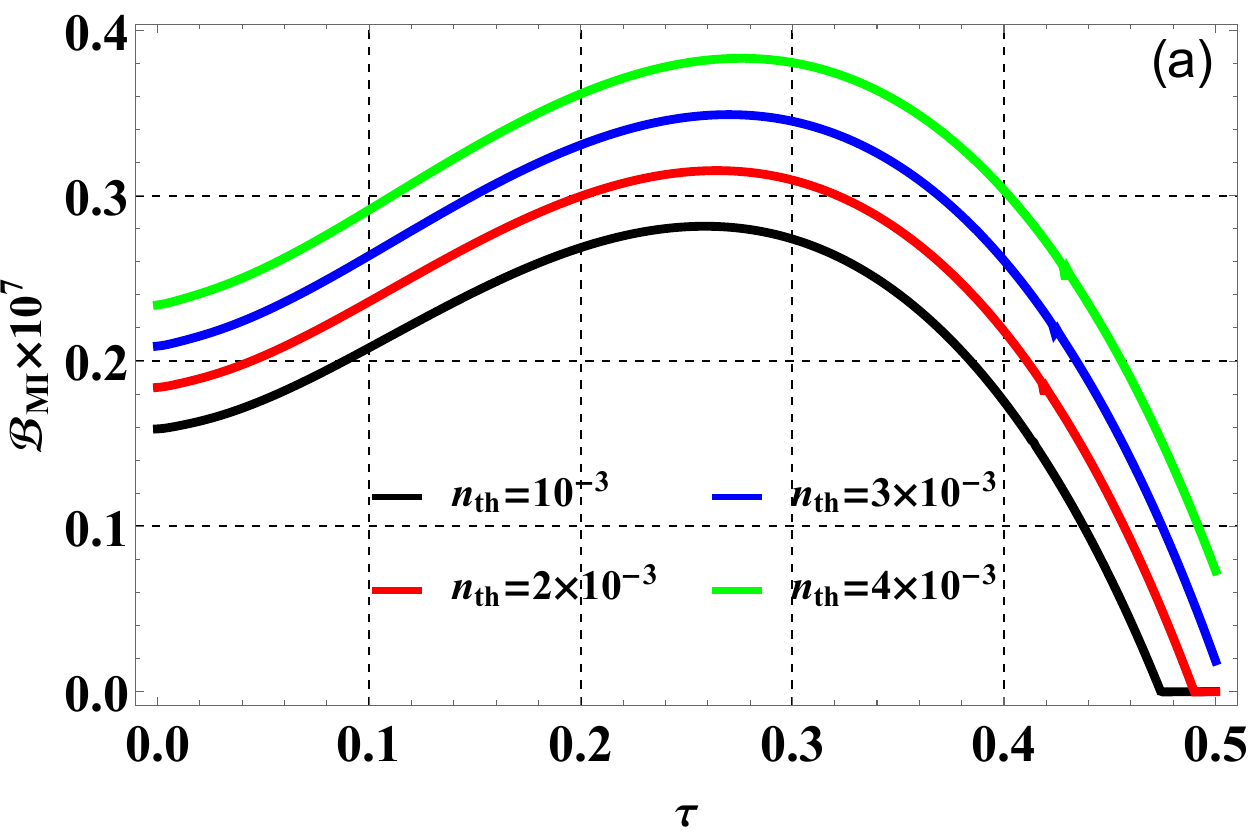} & 
\includegraphics[width=0.32\linewidth]{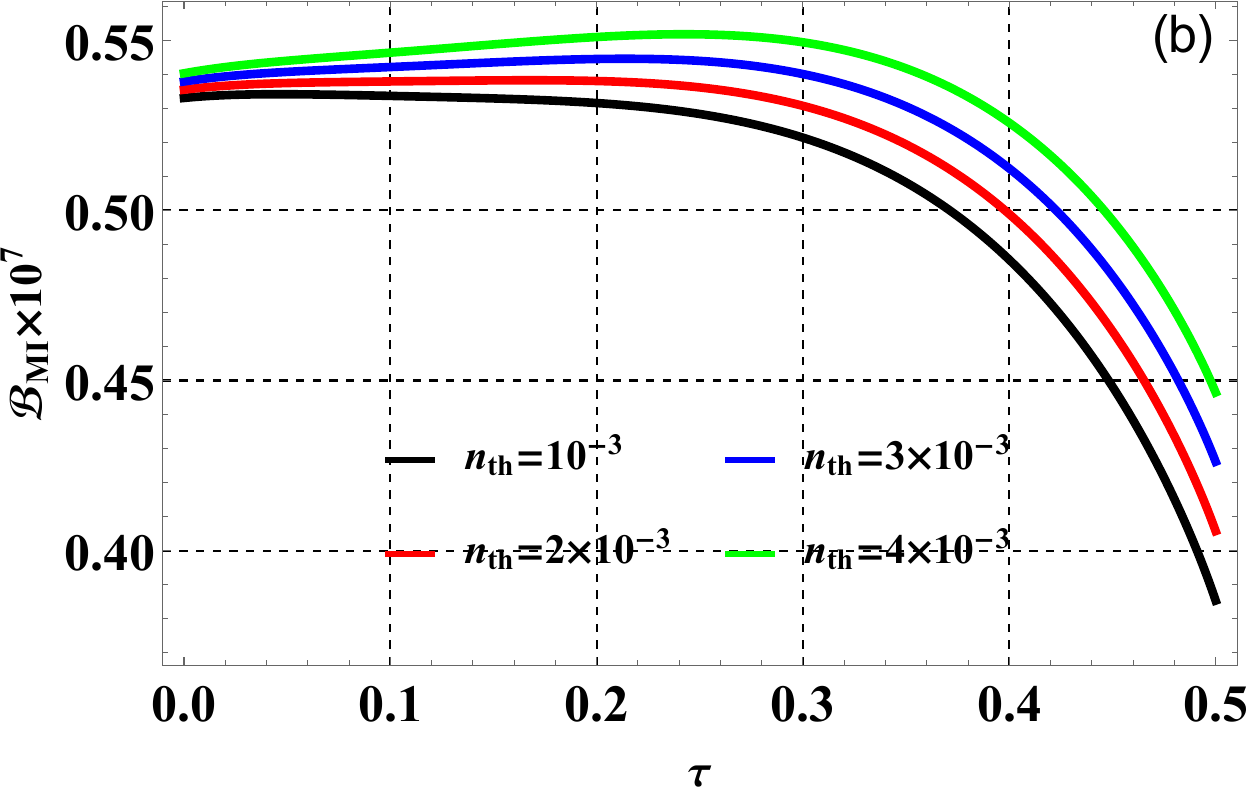} & 
\includegraphics[width=0.32\linewidth]{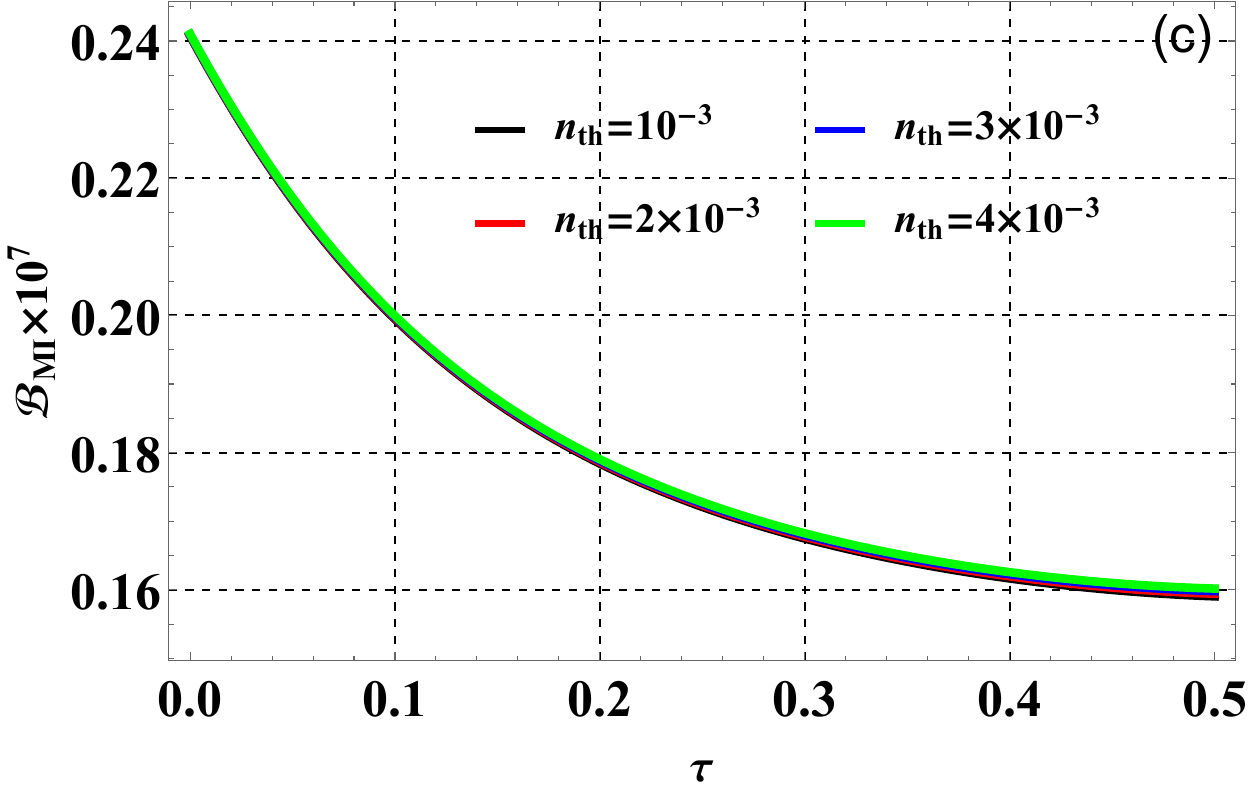} 
\end{tabular}
\caption{Plots of the most informative quantum Cram\'er--Rao bound (QCRB), $\mathcal{B}_{\rm MI}$, as a function of the reflectivity parameter $\tau$ in (a) the absence of parametric amplification ($\lambda=0$) and the squeeze parameter ($\rm r=0.1$), (b) parametric amplification ($\lambda=0.25\kappa_{\rm a}$) and the squeeze parameter ($\rm r=0$), (c) parametric amplification ($\lambda=0.25\kappa_{\rm a}$) and the squeeze parameter ($\rm r=0.1$). for different values of the mean thermal photon number $n_{\rm th}$. The feedback phase is fixed at $\alpha=\pi$ and The PA phase is fixed at$\theta=\frac{\pi}{2}$, while all other system parameters are set according to the values listed in Table~\ref{table}.}
\label{fig12}
\end{figure}

Figure~\ref{fig12} illustrates the dependence of the most informative
quantum Cram\'er--Rao bound (QCRB), $\mathcal{B}_{\rm MI}$, on the
reflectivity parameter $\tau$ of the coherent-feedback beam splitter.
We compare the cases without parametric amplification (PA),
$\lambda=0$, and with PA, $\lambda=0.25\kappa_{\rm a}$, for different
values of the mean thermal photon number $n_{\rm th}$. These results
provide insight into how the coherent-feedback loop, parametric
amplification, squeezing, and thermal fluctuations jointly affect the
ultimate precision of multiparameter quantum estimation. Since
$\mathcal{B}_{\rm MI}$ represents a lower bound on the estimation
variance, a smaller value of $\mathcal{B}_{\rm MI}$ corresponds to a
higher attainable estimation precision.

In Fig.~\ref{fig12}(a), we consider the system in the absence of
parametric amplification, $\lambda=0$, while a finite squeezing
parameter, $\rm r=0.1$, is introduced. For small values of the reflectivity
parameter $\tau$, the most informative QCRB increases with $\tau$.
This behavior corresponds to a degradation of the estimation precision,
since a larger QCRB implies a larger minimum attainable estimation
uncertainty. Physically, in this regime, the coherent-feedback
contribution is relatively weak, and the feedback loop does not yet
provide sufficient control over the intracavity quantum fluctuations
to improve the parameter sensitivity. As $\tau$ increases, the
interference between the directly propagating optical field and the
feedback field becomes more pronounced, thereby modifying the
intracavity field correlations and the information encoded in the
quantum state.

Interestingly, beyond approximately $\tau>0.3$, the behavior is
reversed and $\mathcal{B}_{\rm MI}$ decreases with increasing
reflectivity. Consequently, the estimation precision is enhanced in
this regime. This improvement can be attributed to the stronger
coherent-feedback interaction, which allows the feedback loop to
efficiently reshape the optical fluctuations and enhance the quantum
correlations carrying information about the parameters to be estimated.
The transition around $\tau\simeq0.3$ therefore indicates a crossover
between a regime in which the feedback mainly modifies the estimation
uncertainty and a regime in which it becomes an effective resource for
precision enhancement.

We also observe that $\mathcal{B}_{\rm MI}$ increases with the mean
thermal photon number $n_{\rm th}$. An increase in $n_{\rm th}$ adds
thermal fluctuations to the system and consequently reduces the
distinguishability between quantum states corresponding to different
values of the parameters. From an information-theoretic perspective,
thermal noise decreases the useful quantum information available for
parameter estimation, resulting in a larger QCRB and hence a lower
estimation precision. Thus, the results demonstrate the detrimental
effect of thermal fluctuations on the metrological performance of the
system, while the coherent-feedback mechanism can partially compensate
for this degradation in the appropriate reflectivity regime.

In Fig.~\ref{fig12}(b), we investigate the effect of coherent feedback
in the presence of parametric amplification, with
$\lambda=0.25\kappa_{\rm a}$, while the external squeezing is removed,
$\rm r=0$. In comparison with Fig.~\ref{fig12}(a), the introduction of the
PA substantially modifies the dependence of $\mathcal{B}_{\rm MI}$ on
the reflectivity parameter. In particular, for weak values of $\tau$,
the QCRB exhibits a relatively weak variation as $\tau$ increases,
indicating that the PA partially compensates for the limited
contribution of the coherent-feedback loop in this regime.

The PA modifies the intracavity quantum fluctuations through
phase-sensitive amplification and can redistribute the fluctuations
between the optical quadratures. As a consequence, the information
content of the output quantum state is modified even in the absence of
an externally injected squeezed field. When the reflectivity becomes
larger, particularly for $\tau>0.3$, $\mathcal{B}_{\rm MI}$ decreases
with increasing $\tau$. This decrease directly corresponds to an
enhancement of the estimation precision. The result demonstrates that
the combination of parametric amplification and coherent feedback can
provide a more efficient mechanism for controlling the quantum
fluctuations relevant to parameter estimation. In particular, the
feedback loop can coherently recycle part of the optical field, while
the PA reshapes its quadrature fluctuations, leading to a more
favorable quantum state for simultaneous parameter estimation.

Figure~\ref{fig12}(c) shows the combined action of the parametric
amplifier and the squeezed input field, with
$\lambda=0.25\kappa_{\rm a}$ and $\rm r=0.1$. In this case, the most
informative QCRB decreases continuously with increasing $\tau$ over
the entire investigated range. Therefore, increasing the reflectivity
of the coherent-feedback loop systematically improves the estimation
precision. This behavior is qualitatively different from the cases
shown in Figs.~\ref{fig12}(a) and \ref{fig12}(b), where a crossover
behavior is observed for weak and strong feedback.

The simultaneous presence of external squeezing and parametric
amplification provides two complementary mechanisms for manipulating
quantum fluctuations. The squeezed input field reduces fluctuations
in a selected quadrature, whereas the PA provides an additional
phase-sensitive modification of the intracavity field. The coherent
feedback then controls how these quantum correlations are reinjected
into the system. The combined action of these three resources can
therefore produce a more favorable redistribution of quantum
fluctuations and increase the amount of information accessible for
parameter estimation. This explains the monotonic reduction of
$\mathcal{B}_{\rm MI}$ with $\tau$ observed in Fig.~\ref{fig12}(c).

Another notable feature of Fig.~\ref{fig12}(c) is that the curves
corresponding to different values of $n_{\rm th}$ become nearly
indistinguishable. This indicates that, within the considered
parameter regime, the combined squeezing--PA--feedback configuration
strongly suppresses the sensitivity of the estimation precision to
thermal fluctuations. In other words, the engineered quantum
resources compensate, at least partially, for the information loss
normally caused by thermal noise. This behavior highlights the
potential of coherent feedback assisted by parametric amplification
and squeezing for improving the robustness of quantum metrology in
realistic noisy environments.

Overall, Fig.~\ref{fig12} demonstrates that the reflectivity $\tau$
of the coherent-feedback loop is an important control parameter for
the metrological performance of the system. Without PA, the feedback
effect becomes beneficial mainly beyond a threshold value of
$\tau\simeq0.3$, whereas the inclusion of PA significantly modifies
this dependence. When both parametric amplification and input
squeezing are present, the QCRB is systematically reduced as the
feedback reflectivity increases, leading to a substantial enhancement
of the attainable estimation precision. These results show that
coherent feedback is not merely a passive modification of the optical
configuration; rather, it acts as an active resource for engineering
the quantum correlations and fluctuations that determine the
information available for multiparameter estimation. The combined
use of squeezing, parametric amplification, and coherent feedback
therefore provides an effective strategy for enhancing both the
precision and robustness of quantum parameter estimation in the
presence of thermal noise. The increase in thermal photon number $n_{\rm th}$ generally enhances thermal effect, leading to an increase in $\mathcal{B}_{\rm MI}$ and, consequently, a reduction in estimation precision. However, coherent feedback (CF) and parametric amplification (PA) can mitigate this detrimental effect by controlling the optical fluctuations and enhancing useful quantum correlations. Their combined action reduces the sensitivity of $\mathcal{B}_{\rm MI}$ to thermal noise, resulting in a more robust estimation performance.\\

\begin{figure}[!htb]
\includegraphics[width=0.45\linewidth]{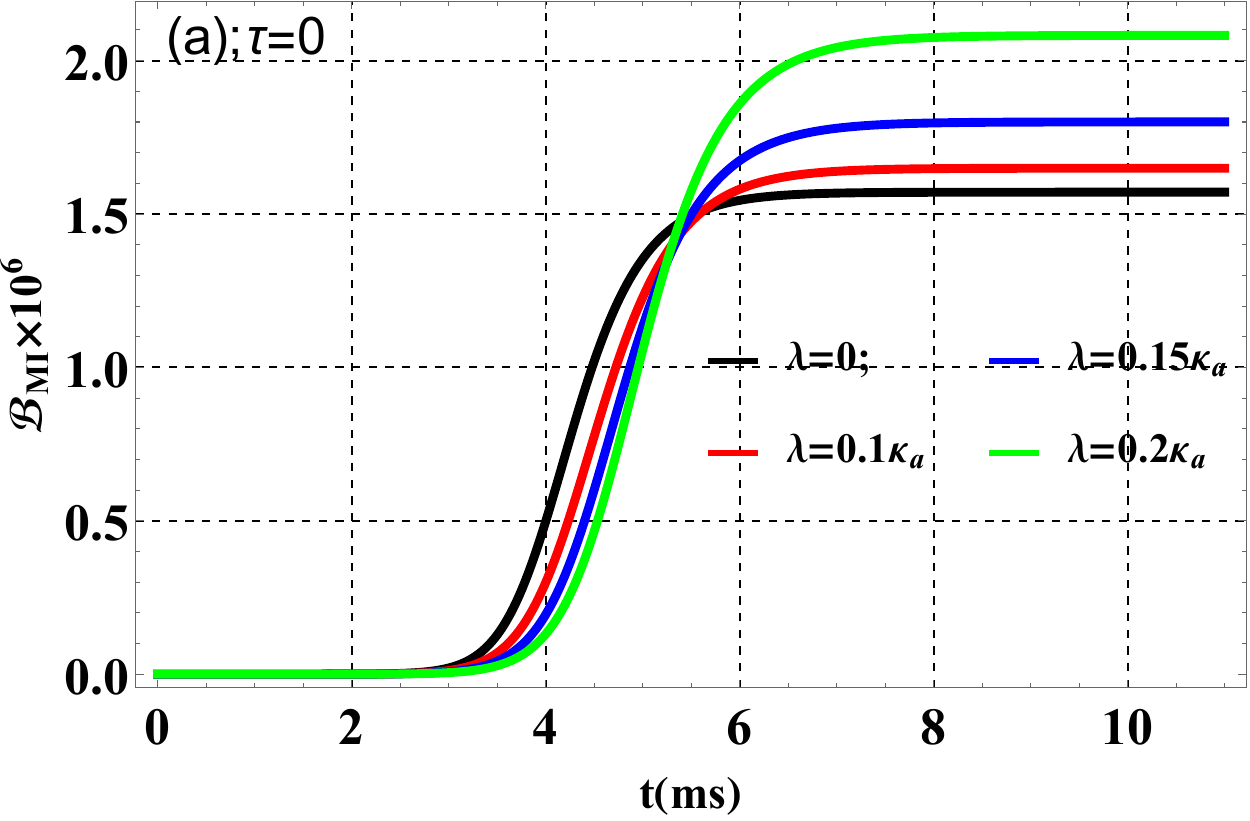}
\includegraphics[width=0.45\linewidth]{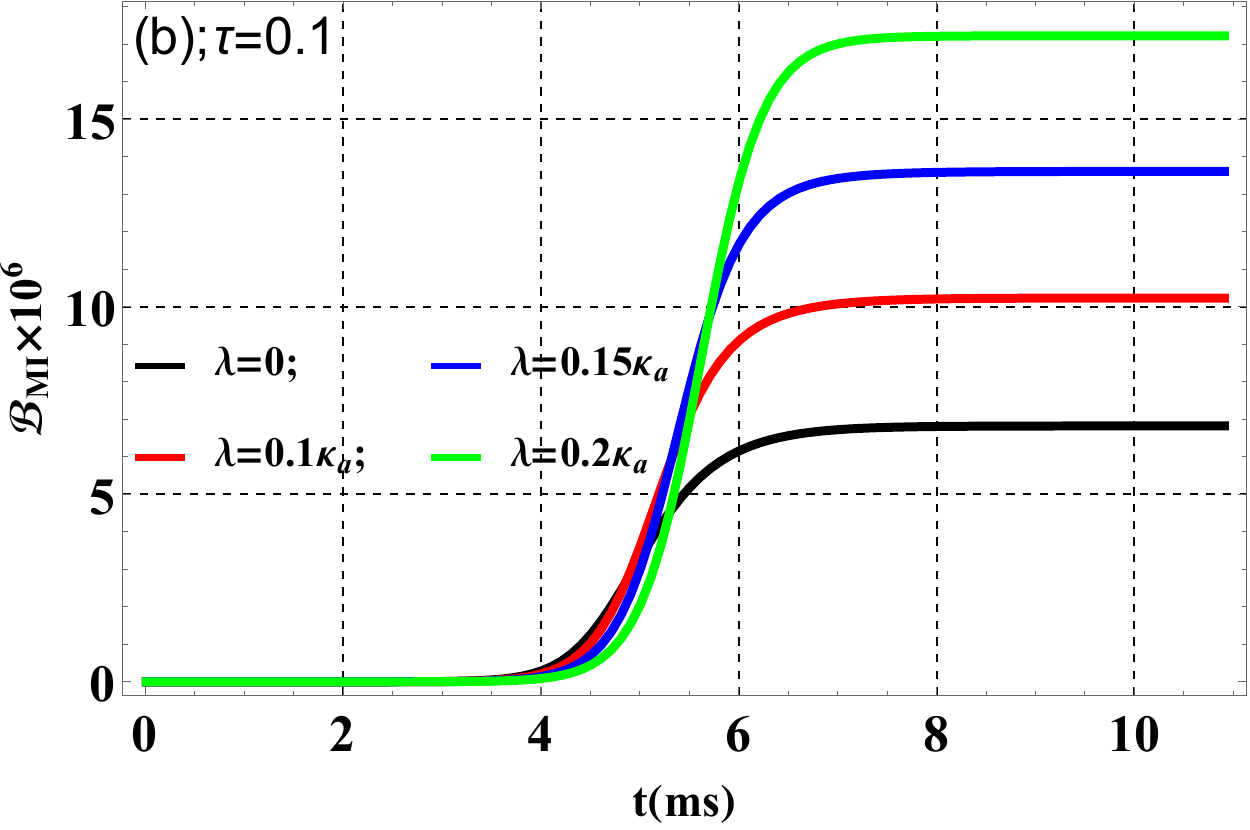}
\caption{Plot of: the evolution of the most informative quantum Cram\'er--Rao bound (QCRB) $\mathcal{B}_{\rm MI}$ (a) in absence of coherent feedback ($\tau=0$) and (b) in presence of coherent feedback ($\tau=0.1$), for different values of the gain parameter $\lambda/\kappa_{\rm a}$, with the phase is fixed at $\theta=\frac{\pi}{2}$ and the feedback phase is fixed at $\alpha=\pi$. The other parameters are provided in the table \ref{table}.}
\label{fig9}
\end{figure}
Figure~\ref{fig9} illustrates the dynamical behavior of the most informative quantum Cram\'er--Rao bound (QCRB), $\mathcal{B}_{\rm MI}$, for different values of the normalized gain parameter $\lambda/\kappa_{\rm a}$, both in the absence of coherent feedback ($\tau=0$) and in the presence of coherent feedback ($\tau=0.1$). This comparison allows us to examine the combined influence of parametric amplification and coherent feedback on the multiparameter estimation performance.

As shown in Fig.~\ref{fig9}(a), corresponding to the case without coherent feedback, the most informative QCRB remains relatively small during the initial stage of the evolution and starts to increase for $\rm t\gtrsim3\mathrm{ms}$. Since $\mathcal{B}_{\rm MI}$ sets a lower bound on the estimation uncertainty, its increase indicates a degradation of the attainable estimation precision at longer evolution times. In the intermediate time interval $3\mathrm{ms}<\rm t<4\mathrm{ms}$, we observe that $\mathcal{B}_{\rm MI}$ decreases as the gain parameter $\lambda/\kappa_{\rm a}$ increases. This reduction of the QCRB demonstrates that, within this dynamical regime, a stronger parametric gain enhances the estimation precision by reducing the minimum achievable uncertainty. However, for longer evolution times, particularly for $\rm t\gtrsim6,\mathrm{ms}$, the behavior is reversed and $\mathcal{B}_{\rm MI}$ increases with increasing $\lambda/\kappa_{\rm a}$. This result indicates that an excessively large gain can become detrimental to the estimation performance at long times, possibly because the amplification of quantum fluctuations is accompanied by an enhancement of the associated noise. Therefore, in the absence of coherent feedback, the gain parameter exhibits a time-dependent role: it can improve the metrological sensitivity within an intermediate temporal regime while degrading the precision at longer times.

Figure~\ref{fig9}(b) shows the corresponding dynamics in the presence of coherent feedback, with $\tau=0.1$. In this case, the feedback loop significantly modifies the temporal evolution of $\mathcal{B}_{\rm MI}$ and leads to an overall improvement of the estimation performance. In particular, around $\rm t\simeq4,\mathrm{ms}$, the most informative QCRB approaches zero, indicating a regime of strongly enhanced estimation precision. Compared with the case without feedback, this result highlights the beneficial role of coherent feedback in controlling the system dynamics and preserving or redistributing the quantum correlations and fluctuations that carry information about the parameters. Moreover, after the initial evolution, the system exhibits a pronounced modification of the QCRB, demonstrating that coherent feedback can substantially improve the metrological response over a specific temporal window. At longer times, however, $\mathcal{B}_{\rm MI}$ increases with increasing $\lambda/\kappa_{\rm a}$, indicating that the enhancement provided by the parametric gain is not unlimited. Although coherent feedback can improve the estimation performance, a large gain may eventually amplify unwanted fluctuations and increase the fundamental estimation uncertainty.

Overall, Fig.~\ref{fig9} reveals the nontrivial interplay between parametric gain and coherent feedback in determining the multiparameter estimation precision. Without coherent feedback, the gain can improve the estimation performance only within a limited temporal region, whereas at longer times a larger gain leads to an increase of $\mathcal{B}_{\rm MI}$ and hence to a reduction of the attainable precision. The introduction of coherent feedback substantially improves the metrological performance and can produce a temporal regime in which the QCRB is strongly suppressed. These results demonstrate that the simultaneous optimization of the gain parameter and the coherent-feedback strength provides an effective strategy for controlling quantum fluctuations and enhancing the precision of multiparameter quantum estimation.

\section{Conclusion} \label{sec7}

In summary, we have theoretically investigated how a coherent feedback loop can improve quantum state transfer from light to matter. Specifically, we have shown that multiparameter quantum estimation of two high quality factor movable mirrors in an optomechanical system comprising two coupled cavities connected via broadband squeezed light can be significantly enhanced by incorporating a coherent feedback loop. Our analysis is conducted in the red sideband regime, where each cavity contains a degenerate parametric amplifier and is driven by a coherent optical pump.\\
Using the quantum Cram\'er–Rao bound as our primary figure of merit, we evaluated the estimation precision through both the symmetric logarithmic derivative (SLD) and right logarithmic derivative (RLD) operators. Our results demonstrate that the presence of the parametric amplifier considerably improves the estimation accuracy by reducing the most informative bound, $\mathcal{B}_{\rm MI}$, through effective control of quantum fluctuations. In addition, the coherent-feedback loop can further reduce $\mathcal{B}_{\rm MI}$ and enhance the multiparameter estimation precision, particularly in the appropriate feedback regime. The combined action of parametric amplification and coherent feedback therefore provides an effective means of improving the overall metrological performance of the system. Despite the detrimental effects of thermal fluctuations, the overall estimation performance remains favorable due to the dominant contributions of parametric amplification and coherent feedback.\\
We further examine the influence of various physical parameters on $\mathcal{B}_{\rm MI}$ in both the dynamical and steady state.
 Additionally, we explored experimentally feasible Gaussian measurement schemes and compared the classical Fisher information for homodyne and heterodyne detection with the SLD quantum Fisher information. Our findings indicate that both detection strategies provide efficient and experimentally accessible routes for parameter estimation in the proposed coupled cavity optomechanical system.

\end{document}